\documentclass[12pt]{article}
\usepackage{putex}
\usepackage{epic}
\begin{document}

\preprint{PUPT-2489}

\title{Evolution of segmented strings}
\authors{Steven S. Gubser}
\institution{PU}{Joseph Henry Laboratories, Princeton University, Princeton, NJ 08544, USA}

\abstract{I explain how to evolve segmented strings in de Sitter and anti-de Sitter spaces of any dimension in terms of forward-directed null displacements.  The evolution is described entirely in terms of discrete hops which do not require a continuum spacetime.  Moreover, the evolution rule is purely algebraic, so it can be defined not only on ordinary real de Sitter and anti-de Sitter, but also on the rational points of the quadratic equations that define these spaces.  For three-dimensional anti-de Sitter space, a simpler evolution rule is possible that descends from the Wess-Zumino-Witten equations of motion.  In this case, one may replace three-dimensional anti-de Sitter space by a non-compact discrete subgroup of $SL(2,{\bf R})$ whose structure is related to the Pell equation.  A discrete version of the BTZ black hole can be constructed as a quotient of this subgroup.  This discrete black hole avoids the firewall paradox by a curious mechanism: even for large black holes, there are no points inside the horizon until one reaches the singularity.}

\date{January 2016}
\maketitle

\tableofcontents

\section{Introduction}
\label{INTRODUCTION}

In \cite{Vegh:2015ska,Callebaut:2015fsa,Vegh:2015yua}, analyses of classical string trajectories were made with the starting point assumption that the string configurations were segmented.  In flat space, this simply means that at a given moment in time, the string is piecewise linear, so the worldsheet is composed of facets each of which is a region in some affine image of ${\bf R}^{1,1}$.  In $AdS_D$, the worldsheet is composed of facets each of which is a region in some $AdS_2$ space of $AdS_D$.  An appealing feature of the analysis in both cases was that the evolution equations are reduced entirely to algebraic operations, first to track the movement of kinks where two string segments meet, and second to determine the outcome of a collision of two kinks.

The purpose of the present work is to simplify the formulation of segmented string motions so that it can easily be applied in $AdS_D$ and $dS_D$ for any $D$ as well as flat space.  I also explore the extent to which we can disregard continuum notions of spacetime and view strings as propagating on spacetimes defined in purely algebraic terms.  The key is to think of the string worldsheet as a collection of spacetime events, namely the collisions of kinks.  On one hand, such an approach may seem antithetical to the common understanding of why string theory works as a quantum theory of gravity---namely that the spatial extent of the string naturally avoids ultraviolet divergences.  On the other hand, finite entropy across unit spacetime area suggests some sort of discreteness in spacetime, and we should therefore be interested in ways in which we can retreat from reliance on a continuous description in formulating string dynamics.

The organization of the paper is as follows.  In section~\ref{FLAT} I introduce the main methods in the simplified context of flat space, where classical string evolution is trivial because one can just add together a left-moving solution and a right-moving solution to the worldsheet wave-equation.  I introduce in section~\ref{CAUSAL} a notion of causal structure motivated by the observation that most of what one needs to know about spacetime in order to talk about segmented strings is the collection of forward-directed null displacements.  Brief discussions of T-duality and worldsheet fermions follow.  In section~\ref{ADS} I explain the evolution of segmented strings in $AdS_D$ and $dS_D$, as well as in some algebraic generalizations of these spacetimes.  In section~\ref{WZW} I explain how fundamental strings moving in $AdS_3$ supported by Neveu-Schwarz three-form flux have a simpler rule for their segmented evolution, relying on the group multiplication law of $SL(2,{\bf R})$.  Because of this reliance, we may consistently demand that all kink collisions occur in some discrete subgroup of $SL(2,{\bf R})$, such as $SL(2,{\bf Z})$.  Taking a stronger view, we can discard $SL(2,{\bf R})$ altogether as a real manifold and think of segmented strings propagating on a discrete spacetime.  Pursuing this line of thought, I develop in section~\ref{BTZ} a discrete analog of the BTZ black hole.  Concluding remarks can be found in section~\ref{CONCLUSIONS}.

\section{Flat spacetime}
\label{FLAT}

All classical results obtained in flat space must be trivial in the sense that they can be recovered from the standard treatment based on left-movers and right-movers.  The putative advantage of the formalism developed here is that it generalizes to at least some curved geometries.  The left-movers / right-movers approach is based on the solution to the string equations of motion:
 \eqn{XSplit}{
  X^\mu(\tau,\sigma) = Y_L^\mu(\sigma^+) + Y_R^\mu(\sigma^-) 
    \qquad\hbox{where}\qquad \sigma^\pm = \tau \pm \sigma \,,
 }
where $\sigma^\pm = \tau \pm \sigma$.  In order to obtain a closed string of finite length, we need $X^\mu(\tau,\sigma)$ to be periodic in the $\sigma$ direction.  The Virasoro constraints are satisfied provided the tangent vectors to $Y_L^\mu(\sigma^+)$ and $Y_R^\mu(\sigma^-)$ (considered as functions of a single variable, $\sigma^+$ or $\sigma^-$) are null.  To obtain a closed string, we need $X^\mu$ to be periodic in the $\sigma$ direction, say with period $\Sigma$.  The periodicity conditions on $Y_L^\mu$ and $Y_R^\mu$ are
 \eqn{YPeriodic}{
  Y_L^\mu(\xi + \Sigma) = Y_L^\mu(\xi) + \pi\alpha' P^\mu \qquad
  Y_R^\mu(\xi - \Sigma) = Y_R^\mu(\xi) - \pi\alpha' P^\mu \,,
 }
and because $Y_L^\mu$ and $Y_R^\mu$ have null tangent vectors, $P^\mu$ must be timelike or null.  Piecewise linear $Y_L^\mu$ and $Y_R^\mu$ lead to segmented string solutions.

All the information about piecewise linear $Y_L^\mu$ and $Y_R^\mu$ is contained in the locations of their turning points: Let's say that within one period of the spatial components, these turning points occur in $Y_L^\mu$ at $\sigma^+_i$ for $i=1,2,\ldots,N_L$, and in $Y_R^\mu$ at $\sigma^-_i$ for $i=1,2,\ldots,N_R$.  We can extend the list of $\sigma^\pm_i$ so that the index $i$ runs across all the integers, and the sequences $\sigma^\pm_i$ stretch all the way into the far past and the far future.  Then these values are the entire list of times when $Y_L^\mu$ or $Y_R^\mu$ has a kink.  Correspondingly, there is a grid of points on the worldsheet, conveniently labeled by $i$ and $j$, where two kinks collide; and this grid is mapped by \eno{XSplit} to a set of points in spacetime, $X_{ij}^\mu$.  From the form of \eno{XSplit}, we know that $X_{i+1,j}^\mu - X_{ij}^\mu$ and $X_{i,j+1}^\mu - X_{ij}^\mu$ are forward-directed null displacements.  The periodicity conditions \eno{YPeriodic} imply
 \eqn{XPeriodic}{
  X_{i+N_L,j}^\mu = X_{ij}^\mu + \pi\alpha' P^\mu \qquad
  X_{i,j+N_R}^\mu = X_{ij}^\mu + \pi\alpha' P^\mu \,.
 }

\subsection{Evolution of serrated slices}
\label{FLATEVOLUTION}

We would like to take a Hamiltonian perspective, where we consider an initial state of the string and then evolve it forward in time.  To this end, it is easiest to start by considering maps from ${\bf Z} \times {\bf Z}$ to the target spacetime, $(i,j) \to X_{ij}$, without any notions of periodicity required for now.  (The index $\mu$ has been suppressed for brevity.)  We are still thinking of $i$ as parametrizing the left-moving light-like direction on the worldsheet, while $j$ parametrizes the right-moving light-like direction.  Now consider traversing ${\bf Z} \times {\bf Z}$ diagonally, where each step in our traverse increments $i$ by $1$ or decrements $j$ by $1$ (but not both).  Let's denote by $S$ a definite choice of how to make this traverse: $S$ then is an ordered subset of ${\bf Z} \times {\bf Z}$ that goes ``all the way across,'' from left to right, in the way we described.  We will make one further stipulation: the range of values of $i$ and $j$ encountered within $S$ must be unbounded in both directions.  In other words, if we project $S$ onto one of the two lightlike directions, it covers all of ${\bf Z}$.  Each segment of $S$ (say from $ij$ to $i+1,j$ or from $ij$ to $i,j-1$) should be thought of as null.  But because we go all the way across the lattice without doubling back, $S$ is effectively a spatial slice of the string worldsheet.  We will describe it as a serrated slice.

Suppose we specify $X_{ij}$ for all $ij \in S$, abiding by the requirement that $X_{i+1,j}-X_{ij}$ is forward-directed null whenever $ij$ and $i+1,j$ are in $S$, and likewise all differences $X_{i,j+1}-X_{ij}$ in $S$ must be forward-directed null.  Then we would like to have a rule for figuring out what all other values of $X_{ij}$ must be.  More particularly, we want a rule that allows us to calculate $X_{ij}$ at a lattice point which is, in an appropriate sense, one step forward from $S$.  To make this precise, consider a point $ij$ on $S$ whose neighboring points are $i+1,j$ and $i,j+1$.  We want to have a rule that specifies $X_{i+1,j+1}$ given $X_{ij}$, $X_{i+1,j}$, and $X_{i,j+1}$.  In flat space, the obvious evolution rule is
 \eqn{FlatSpaceEvolution}{
  X_{i+1,j+1} - X_{ij} = (X_{i+1,j} - X_{ij}) + (X_{i,j+1} - X_{ij}) 
 }
A little thought leads to the conclusion that repetitive use of \eno{FlatSpaceEvolution} allows us to deduce all $X_{ij}$ which can be reached by going forward from $S$: that is, incrementing $i$ and/or $j$.  We could run time evolution the other way by stipulating that if $ij$, $i-1,j$, and $i,j-1$ are all in $S$, then
 \eqn{FlatBackwardEvolution}{
  X_{i-1,j-1} - X_{ij} = (X_{i-1,j} - X_{ij}) + (X_{i,j-1} - X_{ij}) \,.
 }
It is also easy to see that \eno{FlatSpaceEvolution}-\eno{FlatBackwardEvolution} together with the assumption that $S$ goes all the way across ${\bf Z} \times {\bf Z}$ imply
 \eqn{DeltaSoln}{
  X_{i+1,j} - X_{ij} \equiv \Delta_{Li} \qquad\qquad
  X_{i,j+1} - X_{ij} \equiv \Delta_{Rj} \,,
 }
where the $j$-independent quantities $\Delta_{Li}$ are null vectors, as are the $i$-independent quantities $\Delta_{Rj}$.  If we now build sequences $Y_{Li}$ and $Y_{Ri}$ so that
 \eqn{Ysequences}{
  Y_{L,i+1} - Y_{Li} = \Delta_{Li} \qquad\qquad
  Y_{R,i+1} - Y_{Ri} = \Delta_{Ri} \,,
 }
with the added assumption that $X_{00} = Y_{L0} + Y_{R0}$, then we find immediately that
 \eqn{XYrelated}{
  X_{ij} = Y_{Li} + Y_{Rj} \,.
 }
Now let's return to periodicity requirements.  The straightforward notion of periodicity on a serrated slice is to say that $S$ is invariant under the shift $ij \to i-N_L,j+N_R$, where $N_L$ and $N_R$ are positive integers.  We can think of a closed string as a map from a periodic $S$ to spacetime satisfying
 \eqn{SPeriodic}{
  X_{i-N_L,j+N_R} = X_{ij} \qquad\hbox{for all $i,j$.}
 }
Clearly, \eno{SPeriodic} is implied by \eno{XPeriodic}, but not vice versa unless we specify time evolution appropriately.  If we do specify time evolution by \eno{FlatSpaceEvolution}-\eno{FlatBackwardEvolution}, then we are led to the form \eno{XYrelated}, which if plugged into \eno{SPeriodic} can be used to conclude that
 \eqn{YYDiffs}{
  Y_{R,j+N_R} - Y_{Rj} = Y_{Li} - Y_{L,i-N_L} \qquad\hbox{for all $i,j$.}
 }
Denoting the common value of all possible $Y_{R,j+N_R} - Y_{Rj}$ and $Y_{Li} - Y_{L,i-N_L}$ by $\pi\alpha' P$, we immediately recover \eno{XPeriodic}.

\subsection{Examples}

Let's look at some simple examples based on taking $S$ to be the set of all $ij$ such that $i+j$ is either $0$ or $1$.  Choose some null vector $u$, and set $X = 0$ for $i+j=0$ and $X = u$ for $i+j=1$.  This assignment respects the periodicity property \eno{SPeriodic} with $N_L = N_R = 1$.  Then the evolution equations imply for all $ij$ that $X_{ij} = (i+j)u$, and we see that the classical trajectory is collapsed on a null trajectory in the direction of $u$.  The spacetime momentum is $P = u/\pi\alpha'$.

An obvious variant of the previous construction is the ``starburst string,'' where we set $X_{ij}=0$ for $i+j=0$ as before, and $X_{ij} = u_i$ for $i+j=1$ where the $u_i$ are an arbitrary collection of forward-directed null vectors.  Let's focus on the case where the sequence of $u_i$ repeats after $N$ entries.  Then we respect the periodicity property \eno{SPeriodic} with $N_L = N_R = N$.  The solution strategy leading to \eno{XYrelated} can be applied, with the result that $Y_{Ri} = -Y_{L,-i}$ and $Y_{L,i+1} - Y_{Li} = u_{i+1}$.  The overall momentum is $P = {1 \over \pi\alpha'} \sum_{i=1}^N u_i$, which typically is timelike but can be null if the $u_i$ are all parallel.

A special case of the starburst for $N=2$ is the familiar yo-yo string of \cite{Artru:1979ye}.  Most simply, we can work in two dimensions and assume that $u_1 = (\ell,\ell)$ while $u_2 = (\ell,-\ell)$.  Then the string expands from a point at time $t=0$ to a string whose maximal extent at time $t=\ell$ is from $x = -\ell$ to $x = +\ell$.  The string then contracts back to a point at time $t=2\ell$.  A useful check is that the string's energy, $E = 2\ell/\pi\alpha'$, may be calculated either from $P = (u_1 + u_2)/\pi\alpha'$, or by noting that the string's energy comes from its total length of $4\ell$ at full extension.  (Recall that the string doubles back on itself.)

So far we have had in mind that $X_{ij} \in {\bf R}^{d-1,1}$ for some $d$, but one can make different starting assumptions.  For example, we could study strings on a lattice discretization of spacetime by replacing ${\bf R}$ by ${\bf Z}a$ where $a$ is an arbitrary length.  If all the $X_{ij}$ on an initial serrated slice of the discretized string are on the lattice, then all future and past $X_{ij}$ will be too.  Another interesting point is that if we replace ${\bf R}$ by ${\bf Q}a$, where ${\bf Q}$ denotes the rational numbers, then any {\it periodic} string motion (in particular any motion of a finite closed string) that starts out with ${\bf Q}a$-valued coordinates will not only remain ${\bf Q}a$-valued; it will in fact propagate entirely on some integer sublattice.  That is, all coordinate entries will remain in ${\bf Z}\tilde{a}$ for some $\tilde{a}$ commensurate with $a$.  On the other hand, the full set of ${\bf Q}a$-valued segmented string motions is dense in the set of all possible classical string motions on ${\bf R}^{d-1,1}$.  So, in a sense, we can build an understanding all of classical string theory on flat space starting with integer-valued segmented string motions.

\subsection{Causal structure}
\label{CAUSAL}

To generalize further, it is useful to start with some notion of forward-directed null displacements.  Let $M$ be a set of spacetime points, and for each $p \in M$ we associate a subset $N^+(p) \subset M$ comprising all the points of $M$ that can be reached from $p$ by a forward-directed null displacement, and also a subset $T^+(p) \subset M$ of all points that can be reached by a forward-directed timelike displacement.  The minimal requirements that should be obeyed by these subsets are:
 \begin{enumerate}
  \item $N^+(p)$ and $T^+(p)$ are disjoint for any choice of $p \in M$.
  \item $N^+(N^+(p)) \subset N^+(p) \cup T^+(p)$ for any $p \in M$.
 \end{enumerate}
By $N^+(Q)$ for any set $Q \subset M$, we mean the union of $N^+(q)$ for $q \in Q$.  We will refer to $(M,N^+,T^+)$ satisfying the above properties as a weak causal structure on $M$.  Note that we can start from a weak causal structure $(M,N^+,T^+)$ and immediately define a new one $(M,N^-,T^-)$ by saying
 \eqn{NminusDef}{
  N^-(q) = \{ p\colon q \in N^+(p) \} \qquad\qquad
  T^-(q) = \{ p\colon q \in T^+(p) \} \,.
 }
There are some additional properties one might desire, and which are obviously satisfied by null and timelike displacements in ${\bf R}^{d-1,1}$:
 \begin{enumerate}
  \setcounter{enumi}{2}
  \item $p \in N^+(p)$ for any $p \in M$.\label{ZeroElement}
  \item If $q \in N^+(p)$ and $q \neq p$, then $p \not\in N^+(q)$.
  \item $N^+(T^+(p))$, $T^+(N^+(p))$, and $T^+(T^+(p))$ are all subsets of $T^+(p)$ for any $p \in M$.\label{Semigroup}
 \end{enumerate}
With these assumptions in place, one cannot have closed null loops or closed timelike loops, and it follows in particular that $N^+(p) \cap N^-(p) = \{p\}$ and $T^+(p) \cap T^-(p) = \emptyset$.  If a weak causal structure satisfies the additional requirements \ref{ZeroElement} through \ref{Semigroup}, then we characterize it as a strong causal structure.  We could also require:
 \begin{enumerate}
  \setcounter{enumi}{5}
  \item $T^+(p) \subset \bigcup_{n=1}^\infty N^{+n}(p)$ for any $p \in M$, where by $N^{+n}$ we mean the $n$-th iteration $N^+ \circ N^+ \circ \cdots \circ N^+$.\label{GenerateFuture}
  \item If by $N_{\rm all}(p)$ we mean $N^+(p) \cup N^-(p)$, then $M = \bigcup_{n=1}^\infty N_{\rm all}^n(p)$ for any $p \in M$.\label{GenerateAll}
 \end{enumerate}
(The awkward notation $N_{\rm all}(p)$ is due to the fact that we will later want to use $N(p)$ to mean something different in a particular example.)  If a weak causal structure satisfies the additional requirements \ref{GenerateFuture} and \ref{GenerateAll}, then we characterize it as a complete causal structure.

With a weak causal structure in place, we can try to generalize the evolution scheme \eno{FlatSpaceEvolution}.  It's best to start by defining ``forward null triples'' and ``backward null triples'' of spacetime points.  More precisely: We define subsets $F$ and $B$ of the triple product $M \times M \times M$ as follows:
 \eqn{FDef}{
  F &= \{ (X_{00},X_{10},X_{01})\colon X_{00} \in N^-(X_{10}) \cap N^-(X_{01}) \}  \cr
  B &= \{ (X_{11},X_{10},X_{01})\colon X_{11} \in N^+(X_{10}) \cap N^+(X_{01}) \}
 }
Intuitively, an element of $F$---which we refer to as a forward null triple---is the smallest piece of a serrated slice which has enough information to be propagated forward one step, while an element of $B$ (a backward null triple) has just enough information to be propagated backward one step.  For elements $f \in F$ and $b \in B$, we write $f \leftrightarrow b$ to mean that the $X_{10}$ and $X_{01}$ components of $f$ and $b$ are the same.  Now we define an evolution scheme as a map $H$ from a subset of $F$ to a subset of $B$, such that $f \leftrightarrow H(f)$ for all $f$ in the domain of $H$.  For brevity, we could write
 \eqn{HDef}{
  X_{11} = H(X_{00},X_{10},X_{01})
 }
on the understanding that what we really mean is that $H$ maps $(X_{00},X_{10},X_{01}) \in F$ to $(X_{11},X_{10},X_{01}) \in B$.  Initial data can then be specified in the form of a map from a serrated slice $S$ to $M$ such that $X_{i+1,j} \in N^+(X_{ij})$ whenever $i+1,j$ and $ij$ are in $S$, and likewise $X_{i,j+1} \in N^+(X_{ij})$ whenever $i,j+1$ and $ij$ are in $S$.  As before, we evolve forward in time by identifying points $ij \in S$ such that $i+1,j$ and $i,j+1$ are also in $S$, and then $H$ is applied to the triple $(X_{ij},X_{i+1,j},X_{i,j+1})$ to obtain $X_{i+1,j+1}$.  A caveat is that $(X_{ij},X_{i+1,j},X_{i,j+1})$ might not be in the domain of $H$; in such a case, the simplest approach is to declare that the site $ij$ simply doesn't evolve forward.  By construction, $X_{i+1,j+1}$ is in the future of $X_{ij}$, that is $X_{i+1,j+1} \in N^+(X_{ij}) \cup T^+(X_{ij})$.  If the causal structure is strong, then $X_{i'j'}$ is in the future of $X_{ij}$ whenever $i' \geq i$ and $j' \geq j$.  In other words, the evolution of the string is in the future time direction with respect to the causal structure on $M$.

We prefer to consider reversible evolution, namely a function $H$ which is one-to-one, so that an inverse $H^{-1}$ exists from the image of $H$ to its domain.  Then starting with a map from a serrated slice to $M$ (the initial state of the string), one can develop through forward and backward time evolution a unique map from some region in ${\bf Z} \times {\bf Z}$ to $M$.  The region can be thought of as the entire discrete string worldsheet, and it is either the entirety of ${\bf Z} \times {\bf Z}$, or it is bounded in the forward time direction by sites which cannot be evolved further forward, and in the backward time direction by sites which cannot be evolved further backward.

\begin{figure}
  \centerline{\includegraphics[width=5.5in]{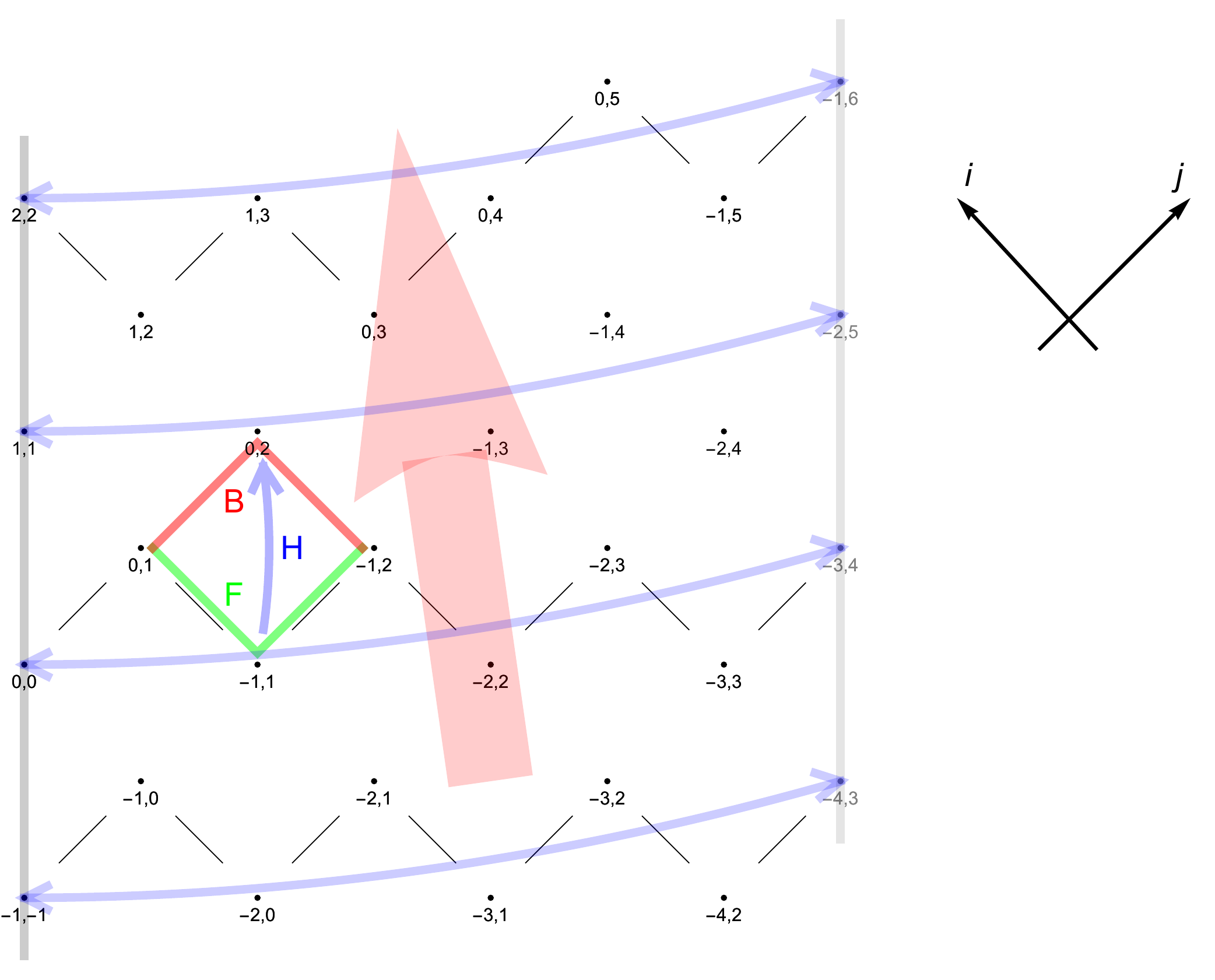}}
  \caption{An example of evolution of serrated slices.  Three serrated slices on the discretized worldsheet are indicated using black line segments from one worldsheet point to the next along the slice.  An initial serrated slice comprises the worldsheet points $(0,0)$, $(0,1)$, $(-1,1)$, $(-1,2)$, $(-2,2)$, $(-2,3)$, $(-3,3)$, and $(-3,4)$ which is identified with $(0,0)$.  Thus $N_L = 3$ and $N_R = 4$.  Similar identifications are made between $(-4,3)$ and $(-1,-1)$, between $(-2,5)$ and $(1,1)$, and between $(-1,6)$ and $(2,2)$, as indicated by the light blue arrows.  The overall direction of forward time evolution is indicated by the wide, light red arrow.  The slice starting at $(-1,-1)$ is the result of seven backward evolution steps from the initial slice.  The slice starting at $(2,2)$ is the result of $12$ forward evolution steps from the initial slice.  The worldsheet representation of a forward null triple is shown in green, and the backward null triple that it gets mapped to by $H$ is shown in red.  More properly, $H$ acts in spacetime to determine the spacetime image of $(0,2)$ once the images of $(-1,1)$, $(0,1)$, and $(-1,2)$ are known.}\label{SeveralSlices}
 \end{figure}

It is straightforward to spell out how one goes from knowledge of an evolution function $H$ to the evolution scheme for serrated slices.  Recall that a serrated slice $S$ includes points $ij$ for all integer values of $i$ and all integer values of $j$.  As a result, $S$ includes infinitely many forward null triples and infinitely many backward null triples.  An easy case to visualize is periodic $S$, with $ij$ identified with $i-N_L,j+N_R$, where $N_L$ and $N_R$ are both necessarily positive.  Then there must be at least one forward null triple and at least one backward null triple in every period.  It is possible but not necessary that $S$ is entirely comprised of overlapping forward and backward null triples.  More likely, there are some forward and backward null triples separated by runs of points all along one lightlike direction on the worldsheet.  In any case, a forward evolution step of a serrated slice consists of choosing one of the forward null triples and mapping it through $H$ to a backward null triple.  (In the case of a periodic serrated slice, of course we mean to map all periodic images of a given forward null triple through $H$ at the same time.)  If there are several forward null triples, it doesn't matter which one is evolved forward first.  Backward evolution proceeds similarly.  Figure~\ref{SeveralSlices} shows an example of the evolution of serrated slices.

\subsection{T-duality}
\label{TDUALITY}

T-duality is a key feature of the theory of free strings.  For simplicity, let's restrict to compactification on a circle of radius $R$.  In the standard approach, spacetime is ${\bf R}^{d-1,1}$ modulo the relation $x \to x + 2\pi R$ where $x$ is one of the spatial coordinates.  Let's replace ${\bf R}$ by ${\bf Z}a$ where $a$ is a length.  Then we need $2\pi R = pa$ where $p$ is some positive integer (not necessarily a prime).  Let us now set $a=1$ for brevity; all the following formulas can be made dimensionally correct by restoring obvious powers of $a$.  

We can present $T$-duality in the framework of discrete spacetime, though with a significant subtlety: We have to work in the ``upstairs'' picture where the compactified extra dimension is regarded as periodic rather than truly compact.  The reason for this is that we will need to distinguish between a purely timelike displacement $(t,x) \to (t+2\pi R,x)$ and a null displacement $(t,x) \to (t+2\pi R,x + 2\pi R)$ that goes once around the circle.  When spacetime is a manifold, it's OK to really identify $x$ and $x + 2\pi R$ because one can tell whether a continuous trajectory is winding around the compact dimension.  Once we have passed to a discrete subset of spacetime points, this distinction is at least much trickier---at least in the discrete topology, it is impossible.

Instead of the periodicity condition \eno{SPeriodic}, we must now require only that the string comes back to itself up to some multiple of the $2\pi R$ displacement in the $x$ direction.  That is,
 \eqn{SigmaSum}{
  \sum_{\alpha \in S} s_\alpha \Delta_\alpha^x = pw \,,
 }
where $w$ is the winding number.  In \eno{SigmaSum} and below, we use $\alpha$ rather than $ij$ to label the null links in a serrated slice $S$, and we set $s_\alpha = 1$ for a link which goes from $ij$ to $i+1,j$ and $s_\alpha = -1$ for a link from $ij$ to $i,j-1$.  By $\Delta_\alpha$ we mean the forward-directed null vector that describes how link $\alpha$ is embedded in spacetime, and by $\Delta_\alpha^x$ we mean its $x$ component.  Thinking in the upstairs picture means that $\Delta_\alpha^x$ is an integer rather than an integer mod $p$, so that we can distinguish between $\Delta_\alpha^x = p$, meaning a displacement once around the circle, and $\Delta_\alpha^x = 0$, which means no displacement in the $x$ direction.  The natural null condition on the $\Delta_\alpha$ is the one that is obvious in the upstairs picture, i.e.~$\Delta_\alpha$ is null in ${\bf Z}^{d-1,1}$.

The spacetime momentum of each link is
 \eqn{SpacetimeMomentum}{
  P_\alpha = {1 \over 2\pi\alpha'} \Delta_\alpha \,,
 }
and so the total momentum in the $x$ direction is
 \eqn{PxValues}{
  P^x = {1 \over 2\pi\alpha'} \sum_\alpha \Delta_\alpha^x = {n \over R} = {2\pi n \over p} \,,
 }
where in the second to last equality we imposed the standard quantum mechanical restriction on possible momenta in terms of an integer $n$.  For this restriction to make sense, we must have
 \eqn{alphaPrimeValue}{
  4\pi^2 \alpha' = pq \,,
 }
where $q$ is also a positive integer.  Then \eno{PxValues} can be rewritten simply as
 \eqn{SumDelta}{
  \sum_{\alpha \in S} \Delta_\alpha^x = qn \,.
 }
The form of \eno{SumDelta} is almost identical to \eno{SigmaSum}.  To perform T-duality, we map
 \eqn{TdualityExplicit}{
  \Delta_\alpha^x \leftrightarrow s_\alpha \Delta_\alpha^x \,,
   \qquad p \leftrightarrow q \qquad w \leftrightarrow n \,.
 }
It is easily checked that $p \leftrightarrow q$ is the usual map $R \to \alpha'/R$.  From a lattice point of view, having exchanged the winding condition with the momentum condition means that the periodicity in the $x$ direction is now $x \sim x + q$ rather than $x \sim x + p$.  Switching the sign on precisely those $\Delta_\alpha^x$ that correspond to links from $ij$ to $i,j-1$ means switching the sign on $\Delta_{Rj}^x$ but not $\Delta_{Lj}^x$.  In other words, we are sending $Y_R \to -Y_R$ while leaving $Y_L$ alone, which is the usual notion of T-duality.  Without referring to $Y_L$ and $Y_R$ explicitly, we could simply note that performing $T$-duality on a serrated slice commutes with the standard evolution \eno{FlatSpaceEvolution}.

We can in principle work in the downstairs picture, where spacetime is ${\bf Z}^{d-2,1} \times {\bf Z}_p$,\footnote{We use ${\bf Z}_p$ to mean the integers mod $p$.} if we declare a displacement $\Delta$ to be null precisely if its projection onto the ${\bf Z}^{d-2,1}$ directions is null.  Development along these lines seems formal in that we are no longer trying to track winding around the ${\bf Z}_p$ direction; however, it is still consistent with the rules of section~\ref{CAUSAL} to use the evolution scheme \eno{FlatSpaceEvolution}.  A large part of what makes T-duality work in the form described in \eno{TdualityExplicit} is that the map $\Delta^x \to -\Delta^x$ is an automorphism of the additive structure on ${\bf Z}$.  We can inquire whether in this downstairs setup some more general automorphism might be used.  An example arises when coordinates are integer-valued and $p = q$, meaning that we are at the self-dual radius; also we require $p$ to be odd.  In the downstairs picture, \eno{SigmaSum} and \eno{SumDelta} read
 \eqn{TwoSums}{
  \sum_{\alpha \in S} s_\alpha \Delta_\alpha^x = 0 \mod p \qquad\qquad
  \sum_{\alpha \in S} \Delta_\alpha^x = 0 \mod p \,.
 }
Because $p$ is odd, these equations can be recast as
 \eqn{CombinedSums}{
  \sum_{\alpha \in S} {1 - s_\alpha \over 2} \Delta_\alpha^x = 0 \mod p \qquad\qquad
  \sum_{\alpha \in S} {1 + s_\alpha \over 2} \Delta_\alpha^x = 0 \mod p \,.
 }
The map $\Delta_\alpha^x \to s_\alpha \Delta_\alpha^x$ means switching the sign of every $\Delta_\alpha^x$ entering into the first equation in \eno{CombinedSums} but not the ones entering into the second equation.  We could instead map all the right-moving displacements through an automorphism $\Sigma$ of the additive structure on ${\bf Z}_p$: $\Delta_\alpha^x \to \Sigma(\Delta_\alpha^x)$ for those $\alpha$ with $s_\alpha = -1$.  An example of such an automorphism is $\Delta_\alpha^x \to r \Delta_\alpha^x$ for $r$ coprime to $p$, and this automorphism provides a candidate generalization of T-duality.  The automorphism is usually not an involution; however, according to Carmichael's theorem, one recovers the identity map after $\lambda(p)$ iterations of $\Sigma$, where $\lambda(p)$ is the so-called reduced totient function.

\subsection{Adding worldsheet fermions}
\label{RNS}

The Ramond-Neveu-Schwarz formalism starts with the following equations of motion for the superstring:
 \eqn{RNSeom}{
  D_{\theta^+} D_{\theta^-} {\bf X}^\mu = 0 \,,
 }
where
 \eqn{Dpm}{
  D_\pm = \partial_{\theta^\pm} + \theta^\pm \partial_{\sigma^\pm} \,.
 }
The bosonic variables $\sigma^\pm = \tau \pm \sigma$ as before, while $\theta^\pm$ are Grassmann variables, and each superfield ${\bf X}^\mu$ is, {\it a priori}, a general function of $\sigma^\pm$ and $\theta^\pm$ (assumed to be smooth).  Without much effort one can show that the general solution of \eno{RNSeom} is
 \eqn{ChiralDecomposition}{
  {\bf X} = {\bf Y}_L(\sigma^+,\theta^+) + {\bf Y}_R(\sigma^-,\theta^-) \,,
 }
and the super-Virasoro constraints are
 \eqn{SuperVirasoro}{
  \partial_{\sigma^+} {\bf Y}_L \cdot D_{\theta^+} {\bf Y}_L = 0 = 
  \partial_{\sigma^-} {\bf Y}_R \cdot D_{\theta^-} {\bf Y}_R \,.
 }
If we expand
 \eqn{YLExpand}{
  {\bf Y}_L = Y_L(\sigma^+) + i \theta^+ \psi_L(\sigma^+) \qquad\qquad
  {\bf Y}_R = Y_R(\sigma^-) + i \theta^- \psi_R(\sigma^-) \,,
 }
then in place of the null-tangent condition on $Y_L(\sigma^+)$ and $Y_R(\sigma^-)$ that we had in the context of the bosonic string, we have from \eno{SuperVirasoro} the more complicated conditions
 \eqn{svComponents}{
  \partial_{\sigma^+} Y_L \cdot \partial_{\sigma^+} Y_L + 
     \psi_L \cdot \partial_{\sigma^+} \psi_L = 0 \qquad\qquad
   \psi_L \cdot \partial_{\sigma_+} Y_L = 0 \,,
 }
and similarly for the right-moving sector.

We would like to distinguish a set of simple solutions \eno{ChiralDecomposition} to the superstring equations of motion which is analogous to the piecewise linear solutions to the bosonic string equations of motion.  Let's examine the problem from the point of view of specifying a certain class of functions ${\bf Y}_L(\sigma^+,\theta^+)$.  Analogous considerations will apply to ${\bf Y}_R$.  An obvious choice is piecewise linear $Y_L(\sigma^+)$ and piecewise constant $\psi_L(\sigma^+)$, where the derivative of $Y_L$ and the value of $\psi_L$ change at the same values $\sigma^+_i$ of the null worldsheet coordinate.  Following our earlier treatment, let's define $Y_{Li} = Y_L(\sigma^+_i)$; also, let's denote by $\psi_{L,i+1/2}$ the constant value of $\psi(\sigma^+)$ for $\sigma^+_i < \sigma^+ < \sigma^+_{i+1}$.  This notation makes clear the point of view that the bosonic data resides at the turning points while the fermionic data resides on the edges connecting them.

Plugging the ansatz just described into \eno{svComponents}, we see from evaluating the first equation away from the special values $\sigma^+_i$ that each linear segment of $Y_L$ must in fact be null: that is,
 \eqn{yCondition}{
  (Y_{L,i+1} - Y_{Li}) \cdot (Y_{L,i+1} - Y_{Li}) = 0 \,.
 }
The condition \eno{yCondition} plus the condition of piecewise linearity implies that the first term in the first equation of \eno{svComponents} vanishes everywhere.\footnote{Precisely at $\sigma^+ = \sigma^+_i$, $\partial_{\sigma^+} Y_L$ changes its value.  However, if we use for $\partial_{\sigma^+} Y_L$ its limiting value as $\sigma^+ \to \sigma^+_i$ from below, then $\partial_{\sigma^+} Y_L \cdot \partial_{\sigma^+} Y_L = 0$ at $\sigma^+ = \sigma^+_i$; likewise if we use the limiting value as $\sigma^+ \to \sigma^+_i$ from above.  It is in this sense that we can say that the first term in the first equation of \eno{svComponents} vanishes everywhere.}  Now consider the second term.  In a neighborhood of $\sigma^+_i$, we see that $\partial_{\sigma^+} \psi_L = (\psi_{L,i+1/2} - \psi_{L,i-1/2}) \delta(\sigma^+ - \sigma^+_i)$ is proportional to a delta-function.  In order for that delta function to drop out of the first equation in \eno{svComponents}, we should require that
 \eqn{psiCondition}{
  \psi_{L,i-1/2} \cdot \psi_{L,i+1/2} = 0 \,.
 }
Note that we are effectively satisfying a stronger constraint than the first equation in \eno{svComponents}, namely $\partial_{\sigma^+} Y_L \cdot \partial_{\sigma^+} Y_L = 0 = \psi_L \cdot \partial_{\sigma^+} \psi_L$.  Passing on to the second equation in \eno{svComponents}, we find by evaluating away from the special points $\sigma^+_i$ that
 \eqn{MixedCondition}{
  \psi_{L,i+1/2} \cdot (Y_{L,i+1} - Y_{Li}) = 0 \,.
 }

We can proceed to a full discretized worldsheet and serrated slices of it by keeping the discussion of the bosonic data exactly the same as it was in the bosonic string case, and adding the fermionic quantities to the links.  To be precise, fermionic variables $\psi_{L,i+1/2,j}$ are located on the links between sites $ij$ and $i+1,j$ for all $i$ and $j$.  We require
 \eqn{SliceConditionsYY}{
  \psi_{L,i-1/2,j} \cdot \psi_{L,i+1/2,j} = 0
 }
if sites $i-1,j$, $ij$, and $i+1,j$ are all present on the serrated slice.  Additionally, we require
 \eqn{SliceConditionsYX}{
  \psi_{L,i+1/2,j} \cdot (X_{i+1,j} - X_{ij}) = 0
 }
if sites $ij$ and $i+1,j$ are present on the serrated slice.  One proceeds analogously with $\psi_R$.  Then the bosonic evolution equation \eno{FlatSpaceEvolution} can be supplemented with the fermionic evolution equation
 \eqn{FermionicAdvance}{
  \psi_{L,i+1/2,j+1} = \psi_{L,i+1/2,j} \qquad\qquad
  \psi_{R,i+1,j+1/2} = \psi_{R,i,j+1/2} \,,
 }
and it is possible to show that the constraints \eno{SliceConditionsYY}-\eno{SliceConditionsYX} are preserved by the evolution.  It is clear from \eno{FermionicAdvance} that $\psi_{L,i+1/2,j} = \psi_{L,i+1/2}$, independent of $j$, and $\psi_{R,i,j+1/2} = \psi_{R,j+1/2}$, independent of $i$.  Technically one might include in the discussion also an auxiliary field $F$, coming from the $\theta^+ \theta^-$ term in the expansion of ${\bf X}$ and defined on the faces of the grid.  This might be of use in less trivial spacetimes; but in flat space, $F$ is immediately set to $0$ through the equations of motion \eno{RNSeom}.

\section{Anti-de Sitter and de Sitter spacetimes}
\label{ADS}

It is natural to start the study of segmented strings in curved spacetime with an examination of $AdS_2$.  In section~\ref{MINIMAL}, starting from $AdS_2$, we will find a simpler expression of the evolution law studied in \cite{Vegh:2015ska,Callebaut:2015fsa,Vegh:2015yua}.  As we will explain, this evolution law can be applied naturally to anti-de Sitter and de Sitter space in any dimension, and to situations where real coordinates are replaced by coordinates valued in some other field, for instance the rationals.  In section~\ref{WZW} we will explain that the evolution law found in section~\ref{MINIMAL} is not the only one we could consider: At least for $AdS_3$, there is another one naturally springing from the fact that $AdS_3$ is the group manifold of $SL(2,{\bf R})$.  This group theoretic evolution law can be extended in an obvious way literally to any group; the main challenge is to decide in a sensible way what group elements generate forward-directed null displacements.

\subsection{$AdS_2$}
\label{MINIMAL}

In \cite{Vegh:2015ska,Callebaut:2015fsa,Vegh:2015yua} it was argued that segmented motions in $AdS_3$ could be composed of worldsheets with facets which are regions of $AdS_2$ subspaces of $AdS_3$.  In this section we ignore higher-dimensional considerations and focus simply on strings on the hyperboloid
 \eqn{AdStwo}{
  -u^2 - v^2 + x^2 = -1
 }
in ${\bf R}^{2,1}$.  Defining the dot product
 \eqn{DotDef}{
  A \cdot B = \eta_{\mu\nu} A^\mu B^\nu \,,
 }
where $\eta_{\mu\nu} = \diag\{-1,-1,1\}$, we can rewrite \eno{AdStwo} as $X \cdot X = -1$.  To define time ordering on the hyperboloid, first define $z = u + i v$.  Then for any two points $X_1$ and $X_2$ on the hyperboloid, form the corresponding $z_1$ and $z_2$, and let $\phi = \arg z_2/z_1 \in (-\pi,\pi]$.  We say that $X_2$ is at a later time than $X_1$ iff $\phi \in (0,\pi)$, and at an earlier time iff $\phi \in (-\pi,0)$.  Evidently, transitivity cannot be relied upon: If $X_2$ is later than $X_1$ and $X_3$ is later than $X_2$, then it may happen that $X_3$ is earlier than $X_1$!  This is the familiar problem of closed timelike curves on the hyperboloid, which is solved by passing to the covering space.  The approach here will be simply to work on the hyperboloid itself.  The presence of closed timelike curves will prevent the causal structure we define from satisfying the fifth condition in section~\ref{CAUSAL}, but we can think of eliminating them in the end by passing to the global cover.  The point of treating the problem directly on the hyperboloid is that we do not actually need to forbid closed timelike curves at the level of the analysis of this section; it is enough that the hyperboloid has a weak causal structure in the sense of section~\ref{CAUSAL}.

To complete the definition of the causal structure, we first note that all the points on the hyperbola which are null separated from a given starting point $X$ have the form $X + \Delta$ where $\Delta \cdot \Delta = 0$ and $X \cdot \Delta = 0$.  Let's define $N^+(X)$ as the set of all $X+\Delta$ which are at a later time than $X$.  Also define $T^+(X) = N^+(N^+(X)) - N^+(X)$, where minus in this context means complement.  It is easily checked that $(M,N^+,T^+)$ indeed defines a weak but complete causal structure, and that this causal structure is unchanged by the $SO(2,1)$ isometries of the hyperboloid.

Let's now ask, what is the most natural evolution function $H$?  By use of an $SO(2,1)$ isometry, we choose the position of $X_{00}$ to be anything we please.  Thus we can assume
 \eqn{abExample}{
  X_{00} = {\small\begin{pmatrix} 1 \\ 0 \\ 0 \end{pmatrix}} ,\quad
   X_{10} = {\small\begin{pmatrix} 1 \\ a \\ \pm a \end{pmatrix}} ,\quad\hbox{and}\quad
   X_{01} = {\small\begin{pmatrix} 1 \\ b \\ \pm b \end{pmatrix}} \,,
 }
where $a$ and $b$ are positive and the two signs can be chosen independently.  Suppose first we choose $-a$ and $+b$.  Then $N^+(X_{10}) \cap N^+(X_{01})$ contains only a single point, namely
\eqn{abResult}{
   X_{11} = {1 \over 1+ab} 
     {\small\begin{pmatrix} 1 - ab \\ a+b \\ -a+b \end{pmatrix}} \,.
 }
It is not hard to check that the generalization of \eno{abExample}-\eno{abResult} which is invariant the action of $SO(2,1)$ is
 \eqn{Evolution}{
  X_{11} &= X_{00} + {\Delta_L + \Delta_R + (\Delta_L \cdot \Delta_R) X_{00} \over
    1 - \Delta_L \cdot \Delta_R / 2}  \cr
   &\hbox{where}\quad
     \Delta_L = X_{10} - X_{00} \quad\hbox{and}\quad \Delta_R = X_{01} - X_{00} \,.
 }
To deduce \eno{Evolution}, we had to assume that the two signs in \eno{abExample} are opposite one another.  If the two signs are the same, then $\Delta_L$ and $\Delta_R$ are parallel, so $\Delta_L \cdot \Delta_R = 0$, and \eno{Evolution} reduces to the abelian evolution law $X_{11} = X_{10} - X_{00} + X_{01}$.  We could in principle consider variants of the evolution law, for example $X_{11} = X_{00} + \lambda(\Delta_L + \Delta_R)$, when $\Delta_L$ and $\Delta_R$ are parallel, but $\lambda \neq 1$ gives rise to problems with energy conservation, and we will not consider it further.  In short, \eno{Evolution} defines the only reasonable evolution law for the hyperboloid.  When formalized into a function $H\colon (X_{00},X_{10},X_{01}) \to (X_{11},X_{10},X_{01})$, one can easily demonstrate that $H$ is a bijection from all of $F$ to all of $B$.  The key step in this demonstration is to show that $\Delta_L \cdot \Delta_R \leq 0$, which is a consequence of $\Delta_L$ and $\Delta_R$ both being forward-directed null displacements.

Without changing any formulas, we can consider another case, namely $dS_2$, which is the same as the hyperboloid \eno{AdStwo} but with a reversal of the identification of time and space, so that the $x$ coordinate is timelike.  $N^+(X)$ should now be the set of all $X + \Delta$ with $\Delta \cdot \Delta = X \cdot \Delta = 0$ where the $x$-component $\Delta^x \geq 0$.  (This is under the convention that increasing $x$ is the future.)  $T^+(X) = N^+(N^+(X)) - N^+(X)$ as before.  \eno{Evolution} can stand as written, but to respect the causal structure of de Sitter, we must require that $\Delta_L$ and $\Delta_R$ should have positive $x$ component.  Now $\Delta_L \cdot \Delta_R \geq 0$, and we are in some danger of making $1 - \Delta_L \cdot \Delta_R / 2$ zero or negative.  Zero is obviously bad because $1 - \Delta_L \cdot \Delta_R / 2$ occurs in the denominator of \eno{Evolution}, but negative is bad too because $X_{11}$ will not be in $T^+(X_{00})$.  All this is in fact a feature not a bug.  Physically, what's going on is that the expansion of de Sitter space drags the string apart, so that only finitely many kink collisions can occur before the string reaches the boundary in the infinite future.  In other words, if $\Delta_L \cdot \Delta_R \geq 2$, then there is no $X_{11}$ because the preceding kinks $X_{10}$ and $X_{01}$ are in different causal patches of de Sitter space, such that $N^+(X_{10}) \cap N^+(X_{01}) = \emptyset$.

\subsection{Algebraic generalizations}
\label{ALGEBRAIC}

With the formula \eno{Evolution} in hand, it is possible to generalize to higher-dimensional spacetimes, or algebraic generalizations of spacetimes.  In place of ${\bf R}^{2,1}$ and $\eta$, consider a vector space $V$ over a field $K$ equipped with a quadratic form $\eta$ that we use as in \eno{DotDef} to define a dot product.  Then the equation $X \cdot X = -1$ defines the algebraic generalization of $AdS_D$ or $dS_D$ that we are interested in, call it $M$.  Assume that the equations $X \cdot X = -1$ and $\Delta \cdot \Delta = 0$ have enough solutions so that we can set up interesting initial conditions on a serrated slice $S$ on the discretized string worldsheet.  Then \eno{Evolution} may be used unchanged.  To go further, let's assume that $K \subset {\bf R}$ so that there is a natural notion of positive and negative, and that $\eta$ has a complete set of eigenvectors over $K$.  If two of the eigenvalues are negative and the rest are positive, then we have a generalized anti-de Sitter spacetime, while if only one is positive and the rest are negative, then we have a generalized de Sitter spacetime.

Causal structure can be defined in analogy to the the approaches for the hyperboloid and $AdS_2$ in section~\ref{MINIMAL}.  For generalized anti-de Sitter spacetimes, we write a general vector $X \in V$ as
 \eqn{Xform}{
  X = u \hat{u} + v \hat{v} + \sum_i x_i \hat{x}_i \,,
 }
where $\hat{u}$ and $\hat{v}$ are the eigenvectors of $\eta$ with negative eigenvalue while $\hat{x}_i$ are the eigenvectors with positive eigenvalue.  Defining $z = u + iv$, one obtains a non-transitive notion of time-ordering---which, as before, can be made transitive by passing to an appropriate covering space.  For generalized de Sitter spacetimes, we write instead
 \eqn{XformDeSitter}{
  X = t \hat{t} + \sum_i x_i \hat{x}_i
 }
where $\hat{t}$ is the unique eigenvector with positive eigenvalue and $\hat{x}_i$ are the other eigenvectors.  A transitive time ordering is then available by ordering with respect to $t$.  In both the anti-de Sitter and de Sitter cases, null displacements on the generalized hyperboloid $X \cdot X = -1$ correspond to null vectors $\Delta$ in $V$ with $X$, and null displacements are forward directed if $X + \Delta$ is later than $X$.  As in the two-dimensional case, we can use $T^+(X) = N^+(N+(X)) - N^+(X)$.  By design, the causal structure is consistent with the evolution rule \eno{Evolution}.  In the anti-de Sitter case, one can again argue that \eno{Evolution} defines a bijection $H\colon F \to B$, while in the de Sitter case it defines a one-to-one map from part of $F$ to part of $B$, as appropriate to the tendency of an inflating spacetime to drag a string apart into separate causal patches.

\subsection{Consistency with equations of motion}

If $K = {\bf R}$ and $\eta = \diag\{-1,-1,1,\ldots,1\}$, then the construction of the previous section gives ordinary $AdS_D$ (up to passing to the global cover, which we will ignore in this section).  If $K = {\bf R}$ and $\eta = \diag\{1,-1,-1,\ldots,-1\}$ then we have ordinary $dS_D$.  The purpose of this section is to argue that the evolution rule \eno{Evolution} then defines a string motion that follows from the Nambu-Goto action.

The main part of the argument is as follows.  Consider $(X_{00},X_{10},X_{01})$ in $F$ such that the image $(X_{11},X_{10},X_{01})$ under $H$ exists, and assume that that $X_{10}-X_{00}$ and $X_{01}-X_{00}$ are not parallel.  Then $X_{00}$, $X_{10}$, and $X_{01}$ together define a plane in ${\bf V}$ that also includes the origin, and intersecting this plane with the hyperboloid $X \cdot X = -1$ gives a copy of $AdS_2$ or $dS_2$.  The point $X_{11}$ as defined by \eno{Evolution} is easily seen to lie in the same plane, so there is a diamond shaped region of $AdS_2$ or $dS_2$ with corners $X_{00}$ in the past, $X_{10}$ and $X_{01}$ on the sides, and $X_{11}$ in the future.  It is consistent with the local equations of motion following from the Nambu-Goto action for the string worldsheet to cover this diamond shaped region.  To see this for the $AdS_D$ case, we need only note that through an $SO(2,D-1)$ transformation we can map the $AdS_2$ subspace to a standard one at $x_2 = x_3 = \ldots = x_{D-1} = 0$.  A string stretched across this subspace is certainly a solution of the Nambu-Goto equations of motion, so at least locally within each diamond shaped region, we are solving the same equations of motion with $SO(2,D-1)$ images of the string stretched across the standard $AdS_2$.  A similar argument can be made in the $dS_D$ case.

From the previous paragraph, the picture we have of segmented strings in $AdS_D$ or $dS_D$ (considered as real manifolds) is that they are assemblies of string segments, each of which is locally static with respect to a suitably defined global time whose choice varies segment by segment.  Within each segment, the string equations of motion are satisfied by the argument of the previous paragraph.  The case where $X_{10}-X_{00}$ and $X_{01}-X_{00}$ are parallel is a degeneration of the non-parallel case, where the $AdS_2$ or $dS_2$ region collapses to a null line along which there can be localized momentum and energy.  It is possible to follow \cite{Ficnar:2013wba} and explicitly modify the Nambu-Goto action by adding terms that support localized null energy-momentum; or one can take the view that motions with localized energy-momentum are best treated as special limits of motions without.  Taking this latter approach, we have only to ask whether the equations of motion are satisfied at each edge, or kink, where one diamond shaped region joins on to another, and whether the equations of motion are satisfied at the corners where kinks collide.

To argue that the equations of motion are satisfied at each edge, it helps to recall that these equations take the form
 \eqn{eomsNG}{
  \partial_a P^a_\mu - \Gamma^\kappa_{\mu\lambda} \partial_a X^\lambda P_\kappa^a = 0
    \qquad\hbox{where}\qquad
  P^a_\mu = -{1 \over 2\pi\alpha'} \sqrt{-h} h^{ab} G_{\mu\nu} \partial_b X^\mu \,.
 }
One matching condition at each edge is that $X^\mu$ is continuous.  Usually at kinks one expects one more matching condition in the form of a first derivative condition, obtained by integrating the equation of motion over a small interval including the kink that joins the two solutions.  Equivalently, we want there to be no delta function contribution, localized at the edge, coming from $\partial_a P^a_\mu$.  The key point is that when this edge is lightlike on the worldsheet, there is no such delta function contribution.  If present, it could only come from a second derivative of $X^\nu$ in $P^a_\mu$.  The second derivative terms, however, are proportional to $\partial_+ \partial_- X^\nu$, where the derivatives are with respect to lightlike worldsheet coordinates $\sigma^\pm$.  If a delta function contribution arose in the equations of motion across an edge at $\sigma^+ = 0$, its coefficient would include a factor
 \eqn{DeltaFactor}{
  \partial_- \left[ X^\nu \Big|_{\sigma^+ = \epsilon} - 
     X^\nu \Big|_{\sigma^+ = -\epsilon} \right] \,,
 }
where $\epsilon$ is a very small positive number.  The quantity in square brackets vanishes as $\epsilon \to 0$ on account of the continuity condition, and the piecewise linearity of $X^\nu$ as a function of $\sigma^\pm$ then implies that the whole expression \eno{DeltaFactor} also vanishes.  So indeed we see that no additional condition is imposed at the first derivative level.

Likewise we claim that there are no additional conditions imposed at a corner where two kinks collide.  The quickest way to argue this is to consider a local description where we focus in on a small neighborhood around a particular corner, call it the $00$ corner.  Then the displacements $\Delta_1 = X_{00} - X_{-1,0}$, $\Delta_2 = X_{00} - X_{0,-1}$, $\Delta_3 = X_{10} - X_{00}$, and $\Delta_4 = X_{01} - X_{00}$ can be viewed as vectors in the tangent space.  Locally the evolution is just like strings in flat space, so if there is a constraint on the four $\Delta_i$ at the corner, it should be present in flat space as well.  But in flat space, these are manifestly independent quantities: In the notation of section~\ref{FLATEVOLUTION}, $\Delta_1 = \Delta_{L,-1}$, $\Delta_2 = \Delta_{R,-1}$, $\Delta_3 = \Delta_{L0}$, and $\Delta_4 = \Delta_{R0}$, and we recall that the $\Delta_{Li}$ and $\Delta_{Rj}$ are all independent of one another.

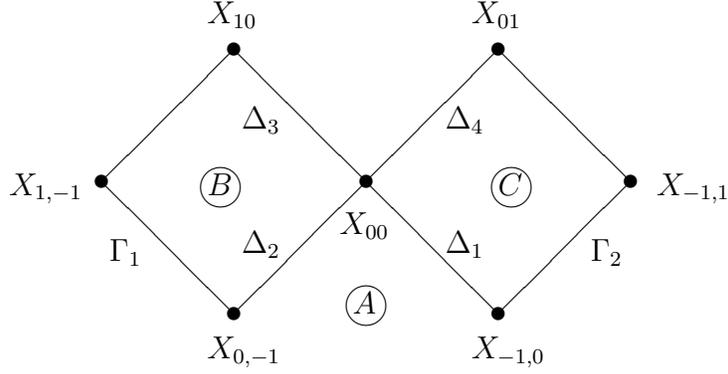
\begin{figure}
  \begin{center}
  \begin{picture}(300,100)(100,40)
   \put(250,100){\circle*{5}}
   \put(200,150){\circle*{5}}
   \put(190,160){$X_{10}$}
   \put(200,50){\circle*{5}}
   \put(190,33){$X_{0,-1}$}
   \put(300,150){\circle*{5}}
   \put(290,160){$X_{01}$}
   \put(300,50){\circle*{5}}
   \put(290,33){$X_{-1,0}$}
   \put(150,100){\circle*{5}}
   \put(115,95){$X_{1,-1}$}
   \put(350,100){\circle*{5}}
   \put(360,95){$X_{-1,1}$}
   \put(250,100){\line(1,1){50}}
   \put(280,120){$\Delta_4$}
   \put(250,100){\line(-1,1){50}}
   \put(203,120){$\Delta_3$}
   \put(250,100){\line(1,-1){50}}
   \put(280,73){$\Delta_1$}
   \put(250,100){\line(-1,-1){50}}
   \put(203,73){$\Delta_2$}
   \put(200,50){\line(-1,1){50}}
   \put(153,70){$\Gamma_1$}
   \put(300,50){\line(1,1){50}}
   \put(335,70){$\Gamma_2$}
   \put(150,100){\line(1,1){50}}
   \put(350,100){\line(-1,1){50}}
   \put(240,80){$X_{00}$}
   \put(195,98){\circle{15}}
   \put(190,95){$B$}
   \put(305,98){\circle{15}}
   \put(300,95){$C$}
   \put(250,53){\circle{15}}
   \put(245,50){$A$}
  \end{picture}
  \end{center}
  \caption{String evolution close to $X_{00}$.}\label{LocalEvolution}
 \end{figure}

To readers familiar with \cite{Vegh:2015ska,Callebaut:2015fsa}, it may seem odd that no conditions are imposed at corners where kinks collide, since in those papers much of the information about string propagation was in fact encoded at the corners.  To see that there is no contradiction, let $\Delta_1$ and $\Delta_2$ be specified, and consider how we gather enough information to determine $\Delta_3$ and $\Delta_4$.  In the approach of the current paper, we must know the triples $(X_{0,-1},X_{1,-1},X_{00})$ and $(X_{-1,0},X_{00},X_{-1,1})$, and from them we deduce $X_{10}$ and $X_{01}$ using \eno{Evolution}.  In other words, we must know $\Gamma_1$ and $\Gamma_2$ as shown in figure~\ref{LocalEvolution} in addition to $\Delta_1$ and $\Delta_2$; then we can get $X_{10}$ and $X_{01}$ from \eno{Evolution} and from them $\Delta_3$ and $\Delta_4$.  In the approach of \cite{Callebaut:2015fsa} (and \cite{Vegh:2015ska} is similar), we require enough local information around the null segment from $X_{-1,0}$ to $X_{00}$ in order to determine the $AdS_2$ regions labeled $A$ and $C$ in figure~\ref{LocalEvolution}, and likewise we require enough information around the null segment from $X_{-1,0}$ to $X_{00}$ in order to determine regions $B$ and $C$.  Then we have enough information to determine the directions of $\Delta_3$ and $\Delta_4$, though not the magnitudes; but that is OK since information from edges further to the left and right will combine with the directions of $\Delta_3$ and $\Delta_4$ to determine $X_{10}$ and $X_{01}$.

\subsection{$AdS_3$ and a discretized WZW model}
\label{WZW}

When we replace ${\bf R}$ by a field $K \subset {\bf R}$, say the rationals, we are giving up most of the points in $AdS_D$ but retaining a dense subset.  An obvious goal is to go further and pass to some discrete set of points distributed more or less uniformly across it, but nowhere densely, similar to the lattice discretizations of flat spacetime.  There is an obvious obstacle to such a step.  Suppose on some $AdS_2$ subspace we are allowed to include initial null displacements $(1,0,0) \to (1,na,-na)$ and/or $(1,0,0) \to (1,nb,nb)$ where $n$ is any positive integer and $a$ and $b$ are fixed and positive.  If we replace $a \to na$ and $b \to nb$ in \eno{abExample}-\eno{abResult}, we see that for $n$ large, there is an accumulation point for $X_{11}$ at $(-1,0,0)$.  This has a perfectly physical interpretation, most easily stated when $a=b$: If we travel a long way toward the boundary of $AdS_2$ as measured by the affine parameter along a null direction, and then turn around and come back the same distance, we will wind up at the same point in space, but approximately half way around the timelike direction on the $AdS_2$ hyperboloid.  This problem has nothing to do with the closed timelike curves on the hyperboloid \eno{AdStwo}; it is instead related to a feature of anti-de Sitter space, namely that there is a definite light-crossing time.  Mathematically, the problem we are seeing has to do with  the presence of $1 - \Delta_L \cdot \Delta_R / 2$ in the denominator of \eno{Evolution}.  When division is part of the evolution law, it is hard to lay down a lattice structure and stay within it starting from generic initial conditions.  So, while it's easy to restrict values of coordinates from ${\bf R}$ down to a field $K \subset {\bf R}$, it is much harder to restrict from ${\bf R}$ to a ring some of whose non-zero elements do not admit multiplicative inverses.

On the other hand, $AdS_3$ is the $SL(2,{\bf R})$ group manifold, and one might naturally expect that segmented strings can propagate so that their kink collisions lie in a discrete subgroup like $SL(2,{\bf Z})$.  This is indeed possible, but not with the evolution law \eno{Evolution}.  A different algebraic evolution law is possible on $AdS_3$; in fact it follows from changing the action from the usual Nambu-Goto action to the WZW action considered for example in \cite{Maldacena:2000hw}.  Then the general solution to the classical equations of motion takes the form
 \eqn{WZWgeneral}{
  X(\tau,\sigma) = Y_L(\sigma^+) Y_R(\sigma^-)
 }
where $X$, $Y_L$, and $Y_R$ all take values in $SL(2,{\bf R})$, and for closed strings we demand
 \eqn{ClosedStringCondition}{
  Y_L(\sigma^+ + \Sigma) = Y_L(\sigma^+) M \qquad\qquad
  Y_R(\sigma^- - \Sigma) = M^{-1} Y_R(\sigma^-)
 }
for some element $M \in SL(2,{\bf R})$.  If we stipulate that there are is no internal CFT, then the Virasoro conditions state that the left- and right-moving currents,
 \eqn{JLR}{
  J_L \equiv (\partial_+ X) X^{-1} \qquad\qquad
  J_R \equiv X^{-1} (\partial_- X) \,,
 }
are null elements of the $SL(2,{\bf R})$ Lie algebra.\footnote{Usually there would be a factor of the level $k$ multiplying $J_L$ and $J_R$, but I suppress it here because it doesn't affect any of the considerations in this section.}  (The $SL(2,{\bf R})$ Lie algebra is composed of traceless $2 \times 2$ matrices.  The null elements are those whose determinant also vanishes.)

The conditions \eno{WZWgeneral}-\eno{JLR} provide a close analogy to the flat space conditions \eno{XSplit}-\eno{YPeriodic} and the condition that the tangent vectors to flat space curves $Y_L^\mu(\sigma^+)$ and $Y_R^\mu(\sigma^-)$ are null.  So it is not surprising that we can proceed to formulate segmented strings on the $SL(2,{\bf R})$ group manifold in fairly close analogy to the flat space case.  In particular, the discrete analogs of $Y_L(\sigma^+)$ and $Y_R(\sigma^-)$ are products of forward-directed null elements of $SL(2,{\bf R})$, by which we mean elements $g \in SL(2,{\bf R})$ with $\tr g = 2$ and $\tr \epsilon g < 0$ where $\epsilon = i\sigma_2$ is the $2 \times 2$ anti-symmetric matrix.  Thus, forward-directed null elements are a special subset of parabolic elements of $SL(2,{\bf R})$.  It can be shown that this notion of forward-directed null displacement on $SL(2,{\bf R})$ coincides with our earlier one if we write 
 \eqn{XgroupForm}{
  X = \begin{pmatrix} u + x & v + y \\ -v + y & u - x \end{pmatrix} = 
    u {\bf 1} + v i\sigma_2 + x \sigma_3 + y \sigma_1 \,.
 }
Indeed, if $g$ is a forward-directed null element of $SL(2,{\bf R})$, then starting at $X$ and moving to $g X$, or more generally $g^n X$ for any $n > 0$, corresponds to moving along a null geodesic (the same geodesic for any $n$) in the natural metric on $SL(2,{\bf R}) = AdS_3$ inherited from ${\bf R}^{2,2}$, and in the positive time direction in the sense of increasing the phase of $u + iv$.  The same is true if one starts at $X$ and moves to $X g$.  Moreover, any forward-directed null displacement can be written {\it either} as left-multiplication or right-multiplication by a forward-directed null element of $SL(2,{\bf R})$.  That is, if $X_1$ is displaced from $X_0$ on $AdS_3$ by a forward-directed null displacement, then $X_1 = g_L X_0$ and $X_1 = X_0 g_R$ where $g_L$ and $g_R$ are forward-directed null elements of $SL(2,{\bf R})$.

Consider now a serrated slice $S$.  The rules for initial data are the same as we have previously used: $X_{i+1,j} - X_{ij}$ must be forward-directed null whenever $ij$ and $i+1,j$ are in $S$ and likewise all differences $X_{i,j+1} - X_{ij}$ in $S$ must be forward-directed null.  To evolve forward in time, in place of \eno{FlatSpaceEvolution} or \eno{Evolution}, we use
 \eqn{WZWevolve}{
  X_{11} = g_L X_{00} g_R \qquad\hbox{where}\qquad
    X_{10} = g_L X_{00} \qquad\hbox{and}\qquad X_{01} = X_{00} g_R \,.
 }
Note that $g_L$ and $g_R$ are forward-directed null elements of $SL(2,{\bf R})$.  Evidently, we can simplify \eno{WZWevolve} to
 \eqn{WZWsimpler}{
  X_{11} = X_{10} X_{00}^{-1} X_{01} \,,
 }
which is a natural generalization of \eno{FlatSpaceEvolution} to $SL(2,{\bf R})$.  To see that \eno{WZWsimpler} is not the same as \eno{Evolution}, it's worth carrying the example \eno{abExample}-\eno{abResult} over to $AdS_3$.  We will quote points on $AdS_3$ in terms of vectors in ${\bf R}^{2,2}$ to facilitate comparison with \eno{Evolution}.  If
 \eqn{WZWabExample}{
  X_{00} = {\small\begin{pmatrix} 1 \\ 0 \\ 0 \\ 0 \end{pmatrix}} ,\quad
   X_{10} = {\small\begin{pmatrix} 1 \\ a \\ -a \\ 0 \end{pmatrix}} ,\quad\hbox{and}\quad
   X_{01} = {\small\begin{pmatrix} 1 \\ b \\ b \\ 0 \end{pmatrix}} \,,
 }
with $a$ and $b$ positive, then according to the group theoretic evolution scheme \eno{WZWsimpler} we have 
 \eqn{WZWabResult}{
   X_{11} =  
     {\small\begin{pmatrix} 1 - 2ab \\ a + b \\ -a + b \\ -2ab \end{pmatrix}} \,;
 }
whereas according to the minimally coupled evolution scheme \eno{Evolution}, choosing $+a$ and $-b$ in \eno{abExample} we would get the result \eno{abResult}, carried over to $AdS_3$ by adding a $0$ entry as the fourth row of the vector.  This is indeed different from \eno{WZWabResult}, and physically the reason is that the WZW model describes a constant field strength $H_3 = dB_2$ which pulls on the string as it moves through $AdS_3$.

The evolution rule \eno{WZWsimpler} could be employed with $SL(2,{\bf R})$ replaced by any group.  If we replace $SL(2,{\bf R})$ with a subgroup, for example $SL(2,{\bf Z})$, then the causal structure discussed above \eno{XgroupForm} can be used to determine what null displacements mean within the subgroup.  In a general setting it is not so simple to give a natural notion of a causal structure.  An obvious class of examples to start with is groups of the form ${\bf R}^{d-1,1} \times H$ where $H$ is some general group which we intend as a replacement for transverse spatial directions (compact or otherwise) and we are considering ${\bf R}^{d-1,1}$ as a group under addition.  The ${\bf R}^{d-1,1}$ factor is equipped with a metric which we write as
 \eqn{FlatMetric}{
  s^2 = -t^2 + \vec{x}^2 \,
 }
Of course, we should be prepared to generalize, for example by replacing ${\bf R}^{d-1,1}$ by a module over a ring and the metric by some appropriate quadratic form.  A trivial but valid version of causal structure is to say that $N^+$ is the direct product of ordinary forward-directed null vectors in ${\bf R}^{d-1,1}$ and all of $H$.  An example of this trivial causal structure came up at the end of the discussion of T-duality in section~\ref{TDUALITY}.  More interesting causal structures could be constructed given a spatial metric on $H$.

\subsection{Minkowski space and the Pell equation}

It was remarked in section~\ref{MINIMAL} that replacing anti-de Sitter space, $AdS_D$, with a level set of a quadratic form on a more general vector space than ${\bf R}^{D-1,2}$ provides an obvious way to generalize the evolution scheme outlined in \eno{Evolution} for classical strings.  In order to set up a particular example, let's first review the connection between the Pell equation and Minkowski space.  I will do that here in more detail than strictly necessary since the subject is appealing on its own.  In two-dimensional Minkowski space, ${\bf R}^{1,1}$, the spacetime interval is
 \eqn{MinkowskiLineElement}{
  s^2 = -c^2 t^2 + x^2 \,,
 }
where $c^2$ is a positive constant.  Let's require that
 \eqn{dDef}{
  d \equiv c^2
 }
is a square-free rational number: that is, $d = a/b$, where $a$ and $b$ are relatively prime and neither is divisible by the square of any integer.  We disallow $d=1$.  The field extension ${\bf Q}[\sqrt{d}]$ comprises all numbers of the form\footnote{Usually one sees field extensions by the square root of an integer, say ${\bf Q}[\sqrt{ab}]$ instead of ${\bf Q}[\sqrt{a/b}]$.  The numbers in these two field extensions are exactly the same since $\sqrt{ab} = b \sqrt{a/b}$; however, the norm defined below is different depending on whether $d = a/b$ or $d = ab$.}
 \eqn{FEelements}{
  z = x + ct
 }
where we commit ourselves to always use the positive square root $c = \sqrt{d} > 0$.  Evidently, ${\bf Q}[\sqrt{d}]$ is a two-dimensional vector space over the rationals, with basis vectors $1 = {\tiny\begin{pmatrix} 1 \\ 0 \end{pmatrix}}$ and $c = {\tiny\begin{pmatrix} 0 \\ 1 \end{pmatrix}}$.  To visualize why ${\bf R}^{1,1}$---or, more precisely, ${\bf Q}^{1,1}$---naturally equates to the field extension ${\bf Q}[\sqrt{d}]$, think of the map $\tiny\begin{pmatrix} t \\ x \end{pmatrix} \to x + ct$ as taking a point in ${\bf Q}^{1,1}$ to an irrational point on the right-moving part of the light-cone.  This is a bijection under our assumption that $d$ is a square free rational number, and we can form another bijection $\tiny\begin{pmatrix} t \\ x \end{pmatrix} \to x - ct$ that we can visualize as projecting onto the left-moving part of the light-cone.\footnote{A related fact is that for square-free $d = c^2$, neither ${\bf Q}^{1,1}$ nor ${\bf Z}^{1,1}$ has any non-trivial points on the lightcone with respect to the Minkowski metric \eno{MinkowskiLineElement}: $N(z) = 0$ implies $z=0$.  By the two-square, three-square, and four-square theorems of Fermat, Legendre, and Lagrange, this situation generalizes, with certain restrictions, to two and three spatial dimensions, but not more.  For example, as a consequence of Legendre's theorem, if $d = c^2 = 7$ (or any positive integer congruent to $7 \mod 8$), then ${\bf Z}^{3,1}$ has no points on the lightcone defined by the interval $s^2 = -c^2 t^2 + \vec{x}^2$.  Could this observation have anything to do with the observed dimensionality of spacetime, or is it only numerology?}

If we write the obvious equality $c (x + ct) = td + cx$ in vector form, it becomes
 \eqn{cMatrix}{
  c \begin{pmatrix} x \\ t \end{pmatrix} = \begin{pmatrix} td \\ x \end{pmatrix} \,.
 }
So, when thinking of the action of ${\bf Q}[\sqrt{d}]$ on itself by left multiplication, we can represent $1$ and $c$ as $2 \times 2$ matrices:
 \eqn{cMatrices}{
  1 = \begin{pmatrix} 1 & 0 \\ 0 & 1 \end{pmatrix} \qquad
  c = \begin{pmatrix} 0 & d \\ 1 & 0 \end{pmatrix} \,.
 }
Thus ${\bf Q}[\sqrt{d}]$ can be represented as all matrices of the form
 \eqn{QdMatrices}{
  z = x + ct = \begin{pmatrix} x & t d \\ t & x \end{pmatrix} \,.
 }
The trace and determinant of the $2 \times 2$ matrix are useful constructions:
 \eqn{TrNorm}{
  \tr(x + ct) = 2x \qquad
  N(x + ct) = -c^2 t^2 + x^2 \,.
 }
The notation $N(x + ct)$ is used because it is the standard mathematical notation for the norm of a field extension.\footnote{In more general field field extensions, $N(\lambda z) = \lambda^n N(z)$ for $\lambda$ in the base field, where the positive integer $n$ may be different from $2$.  For example, $n=4$ for ${\bf Q}[\sqrt[4]{2}]$ because the representation matrices one finds are $4 \times 4$.}  We may treat the complex numbers in the same way, setting $d = -1$; then one has the usual expression $z = x + y\sqrt{-1}$ and the norm $N(z) = x^2 + y^2$.  Pursuing the analogy with complex numbers further, we define
 \eqn{zConjugate}{
  \bar{z} = x - ct \,.
 }
Then $N(z) = z \bar{z}$ is equally valid for $d=-1$ or for positive square-free rational numbers $d$ (and, incidentally, for negative square-free rational $d$).

Rotations in ${\bf C}$ are best represented as $z \to r z$ where $r$ is on the unit circle: $r \bar{r} = 1$.  Generalizing to ${\bf Q}[\sqrt{d}]$, we see that maps $z \to r z$ preserve $N(z)$ precisely when $N(r) = 1$.  These norm-preserving linear maps are boosts.  Explicitly, if we write $r = p + q \sqrt{d}$, then
 \eqn{NrExplicit}{
  N(r) = p^2 - dq^2 = 1 \,.
 }
The relation \eno{NrExplicit} is known as Pell's equation.  A well-known fact (see for example p.~175 of \cite{MillerBook}) is that all solutions of \eno{NrExplicit} with $p$ and $q$ rational take the form
 \eqn{PellForm}{
  r = \pm {1 + \mu/\sqrt{d} \over 1 - \mu/\sqrt{d}} \qquad\hbox{for}\qquad
   \mu \in {\bf Q} \,.
 }
We can describe $\mu$ as the ``half-velocity'' of the boost, in the following sense.  A boost by velocity $v$ can be described as 
 \eqn{StandardBoost}{
  z \to \sqrt{1 + v/c \over 1 - v/c} \, z \qquad\qquad
  \bar{z} \to \sqrt{1 - v/c \over 1 + v/c} \, \bar{z} \,.
 }
(Recall that $z$ and $\bar{z}$ are the light-cone coordinates $x+ct$ and $x-ct$.)  If we define $\mu$ to be the velocity of a boost which must be iterated twice to obtain \eno{StandardBoost}, then we can rewrite \eno{StandardBoost} as
 \eqn{DoubledBoost}{
  z \to {1 + \mu/c \over 1 - \mu/c} \, z \qquad\qquad
  \bar{z} \to {1 - \mu/c \over 1 + \mu/c} \, \bar{z} \,.
 }
Comparing \eno{DoubledBoost} with \eno{PellForm} we see that up to the overall sign on $r$, $\mu$ appearing in \eno{PellForm} indeed has the interpretation of the velocity of a boost which, if iterated twice, leads to the map $z \to r z$.

The original intent of the Pell equation was to restrict $d$, $p$, and $q$ to integers, with $d$ still remaining square-free.  Let's make this restriction now.  Let's also assume for the moment that $p$ and $q$ are in addition positive.  Then there is a fundamental solution $r_1 = p_1 + q_1 \sqrt{d}$ of the Pell equation with the property that all other non-trivial solutions (i.e.~solutions other than $r=1$) take the form $r_n = r_1^n$ for $n=2,3,4,\ldots$.  Furthermore, $p_1/q_1$ is a convergent of the continued fraction expansion of $\sqrt{d}$.  If we drop the restriction that $p$ and $q$ are positive, then the full set of solutions of $N(r) = 1$ is given by 
 \eqn{IntegerPell}{
  r = \pm r_1^n \qquad\hbox{for}\qquad n \in {\bf Z} \,.
 }
This is a special case of Dirichlet's unit theorem, whose statement in this case is that the group of unit elements in ${\bf Z}[\sqrt{d}]$ is generated (up to the sign) by a single element.  (Generally, a unit of a ring is a member of the ring whose multiplicative inverse is also in the ring.)

Thus ${\bf Z}[\sqrt{d}]$ is a lattice version of ${\bf R}^{1,1}$ (technically, it is a free ${\bf Z}$-module spanned by two basis vectors) which is preserved by a group of boosts generated by a ``fundamental boost'' $r_1$.  If we write a general boost as
 \eqn{GeneralBoost}{
  \begin{pmatrix} x' \\ t' \end{pmatrix} = 
    r \begin{pmatrix} x \\ t \end{pmatrix} \qquad\hbox{where}\qquad
  r = \gamma \begin{pmatrix} 1 & 0 \\ 0 & 1 \end{pmatrix} + 
      {\gamma v \over c^2} \begin{pmatrix} 0 & c^2 \\ 1 & 0 \end{pmatrix} 
        \quad\hbox{and}\quad
      \gamma = {1 \over \sqrt{1-v^2/c^2}} \,,
 }
then $r_1$ has the special property that its $\gamma$ and $\gamma v/c^2$ are integers, and any lattice-preserving boost whose $2 \times 2$ matrix form has integer entries can be reached, up to a sign, by applying the boost $r_1$ or its inverse some finite number of times.  Evidently, this situation is more interesting than rotations of ${\bf Z}[\sqrt{-1}]$, which are all obtained by multiplying by a power of $\sqrt{-1}$.

\subsection{A discrete version of the BTZ black hole}
\label{BTZ}

The BTZ black hole is locally $AdS_3$, but with identification of points related by a discrete group isomorphic to ${\bf Z}$.  Specifically, we start with $AdS_3$ defined as the locus of points in ${\bf R}^{2,2}$ satisfying
 \eqn{AdSWithC}{
  -c^2 (u^2+v^2) + x^2 + y^2 = -\ell^2 \,.
 }
We form
 \eqn{zwDefs}{
  z = u + {x \over c} \qquad\qquad
  \bar{z} = u - {x \over c} \qquad\qquad
  w = v + {y \over c} \qquad\qquad
  \bar{w} = v - {y \over c} \,.
 }
It will soon become clear why \eno{zwDefs} works better in the current context than the definition \eno{FEelements} which we used in the previous section.  Approximately following \cite{Banados:1992gq}, away from extremality we construct the BTZ black hole by identifying
 \eqn{zwIdent}{
  z \sim e^{2\pi r_+/\ell} z \qquad\qquad
  \bar{z} \sim e^{-2\pi r_+/\ell} \bar{z} \qquad\qquad
  w \sim e^{2\pi r_-/\ell} w \qquad\qquad
  \bar{w} \sim e^{-2\pi r_-/\ell} \bar{w} \,,
 }
where $r_\pm$ are the outer and inner horizon radii, satisfying $0 \leq r_- < r_+$.  The resulting black hole thermodynamics (again following \cite{Banados:1992gq}) is
 \eqn{MandJ}{\seqalign{\span\TL & \span\TR &\qquad\qquad \span\TL & \span\TR}{
  M &= {c_{\rm Vir} \over 12} {\hbar \over c\ell} 
    {r_+^2 + r_-^2 \over \ell^2} &
  & J = {c_{\rm Vir} \over 12} \hbar {r_+ r_- \over \ell^2}  \cr
  S &= {\pi c_{\rm Vir} \over 3} {r_+ \over \ell} &
  & T_+ \pm T_- = {\hbar c \over \pi\ell^2} r_\pm
 }}
where
 \eqn{cVirasoro}{
  c_{\rm Vir} = {c^3 \over \hbar} {3\ell \over 2G_3}
 }
is the central charge $c_{\rm Vir}$ of the Virasoro algebra of asymptotic symmetry generators.

To pass to a discrete version of BTZ, consider ${\bf Z}^{2,2}$ with coordinates $(u,v,x,y)$ equipped with the quadratic form
 \eqn{ZttMetric}{
  s^2 = -u^2 - v^2 + {1 \over c^2} (x^2 + y^2) \qquad\hbox{where}\qquad
    {1 \over c^2} = d
 }
and $d$ is a square-free integer.  Setting $c^2 = 1/d$ contrasts with the discussion of the previous section but is a better choice here, as we will see momentarily.  (The interval \eno{ZttMetric} differs by a factor of $c^2$ from the left hand side of \eno{AdSWithC} because I wanted the metric \eno{ZttMetric} to be expressible wholly in terms of ring operations on ${\bf Z}$.)  In terms of the variables $z$ and $w$ defined in \eno{zwDefs}, the interval \eno{ZttMetric} can be expressed compactly as $s^2 = -N(z) - N(w)$.  Next define
 \eqn{gFromzw}{
  X = \begin{pmatrix} z & w \\ -\bar{w} & \bar{z} \end{pmatrix} \,,
 }
and note that $s^2 = -\det X$.  If $\ell = c$, the equation \eno{AdSWithC} can be rewritten as
 \eqn{AdSAgain}{
  \det X = N(z) + N(w) = 1 \,.
 }
Matrices of the form \eno{gFromzw} (with integer $u$, $v$, $x$, and $y$) satisfying \eno{AdSAgain} form a group under matrix multiplication.  This group can be described as $SL(2,{\bf Z})_d$; or it can be thought of as $U(2)_d$ in the sense that its elements satisfy $X^{-1} = X^\dagger$ where $\dagger$ means matrix transposition combined with conjugation, $z \to \bar{z}$ and $w \to \bar{w}$.  If we allow $d=1$, then we do not quite recover $SL(2,{\bf Z})$; instead, $SL(2,{\bf Z})_1$ is the subgroup of $SL(2,{\bf Z})$ whose elements $X = \tiny\begin{pmatrix} A & B \\ C & D \end{pmatrix}$ have $A-D$ and $B-C$ both even.  In any case, $SL(2,{\bf Z})_d$ is a subgroup of $SL(2,{\bf R})$, so it inherits the causal structure introduced above \eno{XgroupForm}.

To connect smoothly with the previous section, one might have expected the definition $z = x + u \sqrt{d}$ in place of $z = u + x \sqrt{d}$.  The trouble is, if we set $z = x + u \sqrt{d}$, then the condition $\det X = 1$ that is preserved under group multiplication defines $AdS_3$ with signature $+$$-$$-$, where $x$ and $y$ combine to give the compact timelike direction while $u$ and $v$ are spacelike.  Swapping $(x,y)$ for $(u,v)$, we arrive back at the construction of the previous paragraphs, starting with \eno{AdSWithC} and \eno{zwDefs}.  In other words, $c^2 = 1/d$ is actually the only choice we can make consistent with the group structure we are interested in.

The identification \eno{zwIdent} that defines the BTZ black hole can be rewritten as
 \eqn{BTZid}{
  X \sim \tilde{X} \equiv \begin{pmatrix} r_z z & r_w w \\ 
    -r_w^{-1} \bar{w} & r_z^{-1} \bar{z} \end{pmatrix}
 }
where 
 \eqn{rzrw}{
  r_z = p_z + q_z \sqrt{d} = r_1^{n_z} \qquad\qquad\hbox{and}\qquad\qquad
  r_w = p_w + q_w \sqrt{d} = r_1^{n_w}
 }
are fixed elements of ${\bf Z}[\sqrt{d}]$ with $N(r_z) = N(r_w) = 1$.  We also require that the rational part of $r_z$ and $r_w$ is positive.  This last requirement is the moral equivalent of stipulating that these boosts should be continuously connected to the identity.  It implies that if $r_1$ is the fundamental solution of $N(r) = 1$ over ${\bf Z}[\sqrt{d}]$, then $r_z = r_1^{n_z}$ and $r_w = r_1^{n_w}$ for $n_z$ and $n_w$ both integer.  For consistency with the identifications \eno{zwIdent} we further require non-negative $n_z$ and $n_w$.  Setting $d=1$ at this point is a problem, because then no choice of $r_z$ and $r_w$ other than $r_z=r_w = 1$ is allowed.  To see this, note that there are elements of $SL(2,{\bf Z})_1$ in which any desired one of $z$, $\bar{z}$, $w$, or $\bar{w}$ equals $1$.  Then for $\tilde{X}$ to be also in $SL(2,{\bf Z})_1$, we need $r_z$, $r_z^{-1}$, $r_w$, and $r_w^{-1}$ all to be integers; since we restrict $r_z$ and $r_w$ to be non-negative, they must therefore be $1$.  We avoid this overly restrictive situation by insisting that $d>1$ is a square-free integer.

Comparing the identifications \eno{zwIdent} and \eno{BTZid}, we see that they match provided
 \eqn{IdentMatch}{
  {r_+ \over \ell} = {1 \over 2\pi} \log r_z \qquad\qquad
  {r_- \over \ell} = {1 \over 2\pi} \log r_w \,.
 }
Plugging into the formula \eno{MandJ} for the entropy and using \eno{rzrw}, one finds
 \eqn{QuantizedEntropy}{
  S = {c_{\rm Vir} \over 6} n_z \log r_1 \,.
 }
This formula is striking not only because the permitted values of $S$ are evenly spaced, but because the spacing depends on $d$ in an irregular fashion.  For $d=3$, one finds $r_1 = 2 + \sqrt{3}$, and $\log r_1 \approx 1.32$; but for $d=61$ (famously), one finds the numerically large result $r_1 = 1766319049 + 226153980 \sqrt{61}$, such that $\log r_1 \approx 22.0$.  Physically, a large $\log r_1$ means that the smallest black hole (above the zero-size black hole associated with the RR vacuum) is very large compared to the $AdS_3$ radius.  Specifically, the horizon area (really a length) scales as $A \propto \ell n_z \log r_1$.

One can also obtain expressions for the dimensionful quantities $M$, $J$, and $T_\pm$; in so doing, it should be noted that the requirement $\ell = c$, which follows from setting $\det X = 1$, is essentially a choice of units.  In particular, we are {\it not} restricting attention to Planck-scale $AdS_3$ because the Planck scale remains free even after we choose units.  The dual statement in field theory is that we have not fixed $c_{\rm Vir}$.  

 \begin{figure}
  \centerline{\includegraphics[width=6.5in]{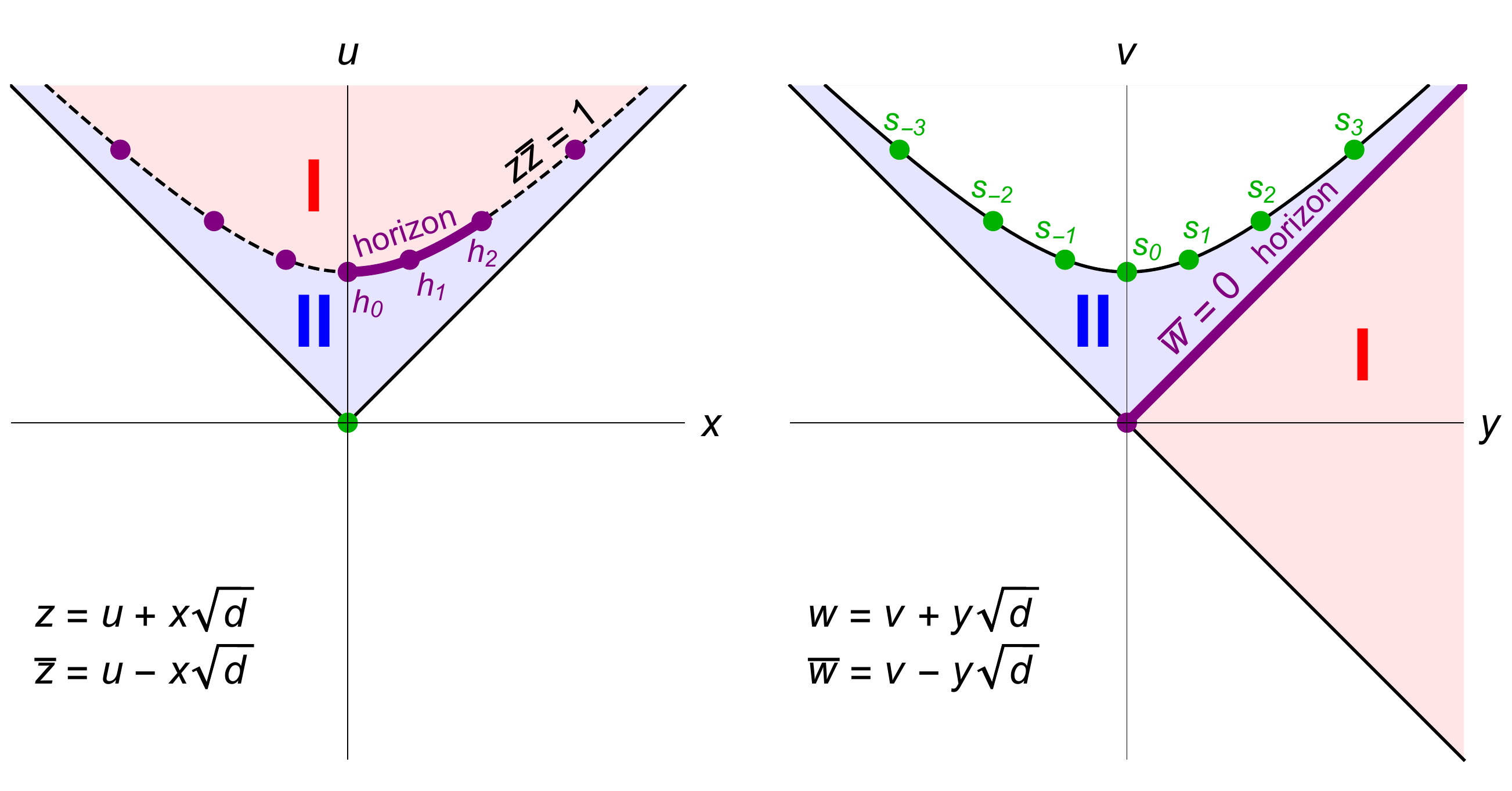}}
  \caption{The outer region {\bf I} and inner region {\bf II} of the BTZ black hole, projected onto the $z$-$\bar{z}$ and $w$-$\bar{w}$ planes.  The points $s_n$ of \eno{snDef} are shown as green dots, while the points $h_n$ of \eno{HorizonPoints} are shown as dots along the horizon $N(z) \equiv z\bar{z} = 1$.  The particular case shown here has an identification with $n_z = 2$ and $n_w = 0$, corresponding to an uncharged black hole.  There are two distinct points in $SL(2,{\bf Z})_d$ on the horizon, namely $h_0$ and $h_1$, and there are no points in $SL(2,{\bf Z})_d$ in the interior of region ${\bf II}$.}\label{BTZfig}
 \end{figure}

The full geometry of the BTZ construction is somewhat complicated, but the main features that make it a black hole can be understood by focusing on two regions:
 \eqn{TwoRegions}{\seqalign{\span\TT &\qquad \span\TL & \span\TR\,, &\quad \span\TL & \span\TR\,, &\quad \span\TL & \span\TR\,, &\quad \span\TL & \span\TR}{
  {\bf I} (outside): & N(z) &> 1 & z+\bar{z} &> 0 & w &> 0 & \bar{w} &< 0 \cr
  {\bf II} (inside): & 1 > N(z) &> 0 & z+\bar{z} &> 0 & w &> 0 & \bar{w} &> 0 \,.
 }}
For now, let's regard these regions as defined in the real manifold $AdS_3$, with real-valued global coordinates coordinates $z$, $\bar{z}$, $w$, and $\bar{w}$ as introduced in \eno{zwDefs}.  The boundary between {\bf I} and {\bf II} is the black hole horizon, in the sense that if $X \in {\bf I}$ and $Y \in {\bf II}$, then it is possible to have $Y \in N^+(X)$, but one can never have $X \in N^+(Y)$.  This is with the notion of causality inherited from $AdS_3 = SL(2,{\bf R})$, namely that $N^+(X)$ is all elements of $SL(2,{\bf R})$ of the form $pX$ where $p$ is a forward-directed null element of $SL(2,{\bf R})_d$.  The horizon is a two-dimensional submanifold of $SL(2,{\bf R})$ with one direction spacelike and the other null.  The identification $X \to \tilde{X}$ preserves regions ${\bf I}$ and ${\bf II}$ as well as the horizon between them.  Region ${\bf I}$ becomes an outside region of the BTZ black hole, while region ${\bf II}$ is an inside region.

Region ${\bf II}$ is empty when we restrict to $SL(2,{\bf Z})_d$.  The easiest way to see this is that $N(z) = u^2 - d x^2$ is an integer, so it is impossible to satisfy the constraint $1 > N(z) > 0$ of \eno{TwoRegions}.  Suppose we expand region ${\bf II}$ to include points with $N(z) = 0$, which immediately implies $N(w) = 1$.  There is an infinite sequence of such points, namely 
 \eqn{snDef}{
  s_n = \begin{pmatrix} 0 & r_1^n \\ -r_1^{-n} & 0 \end{pmatrix}
     \qquad\hbox{for}\quad n \in {\bf Z} \,.
 }
The identification map $X \to \tilde{X}$ as defined in \eno{BTZid} maps each $s_n$ to itself if $n_w = 0$ (the uncharged BTZ black hole), whereas $s_n \to s_{n+n_w}$ for the charged case.  Let's focus on the uncharged case for simplicity.  In the geometry of the uncharged BTZ black hole as a real manifold, the locus of fixed points, $z = \bar{z} = 0$ with $N(w) = 1$, is the singularity inside the black hole.  Using the explicit coordinate maps of \cite{Banados:1992gq}, this locus of points corresponds to $r=0$ in the usual parametrization of the uncharged BTZ black hole as
 \eqn{UnchargedMetric}{
  ds^2 = -{r^2 - r_+^2 \over \ell^2} c^2 dt^2 + {\ell^2 \over r^2 - r_+^2} dr^2 + 
    r^2 d\phi^2 \,.
 }
Thus the points $s_n$ are all at $r=0$.  In summary: in the discrete construction, there is nothing inside the uncharged black hole except the singularity, which resolves into an infinite series of fixed points $s_n$ of the isometry \eno{BTZid}.

A similar discussion can be carried out for the horizon.  With the straightforward definition of the horizon between ${\bf I}$ and ${\bf II}$ as the set of points with $N(z) = 1$, $z + \bar{z} > 0$, $\bar{w} = 0$, and $w > 0$, we find immediately that there are no points in $SL(2,{\bf Z})_d$ which are on the horizon.  That is because $\bar{w} = 0$ implies $w=0$ in ${\bf Z}[\sqrt{d}]$.  If we relax the definition slightly by requiring $w \geq 0$ instead of $w > 0$, then the points in $SL(2,{\bf Z})_d$ on the horizon are
 \eqn{HorizonPoints}{
  h_n \equiv \begin{pmatrix} r_1^n & 0 \\ 0 & r_1^{-n} \end{pmatrix} \qquad\hbox{for}\quad
   n \in {\bf Z} \,.
 }
The identification map \eno{BTZid} sends $h_n \to h_{n+n_z}$, so $n_z$ of these points remain distinct.  (This is equally true for the uncharged and charged cases.)  Thus we have an appealing picture of a horizon comprising $n_z$ points and carrying an entropy proportional to $n_z$.  It is as if there are
 \eqn{Wone}{
  W_1 \equiv r_1^{c_{\rm Vir}/6}
 }
distinct microstates per horizon point, with the choice of microstate at each point being independent of one another.  $W_1$ as defined in \eno{Wone} has no reason to be an integer, or even rational; but possibly there is some sense in which it may be used to capture an asymptotic count of the number of states.

The BTZ construction applied to $SL(2,{\bf Z})_d$ amounts to a sort of lattice in $AdS_3$ (namely the points of $SL(2,{\bf Z})_d$ itself) which respects the identifications \eno{BTZid} used to construct BTZ black holes of particular masses and angular momentum.  But in light of the discussion leading to \eno{WZWsimpler}, it should be something more, namely a non-compact discrete set on which strings can propagate consistently, with causal properties inherited from the BTZ black hole over the reals.  To see that string propagation on our quotient space is consistent, first note that we can express \eno{BTZid} as
 \eqn{idGroup}{
  X \sim \tilde{X} = g_L X g_R^{-1} \qquad\qquad\hbox{where}\qquad\qquad
   g_L = \begin{pmatrix} r_L & 0 \\ 0 & r_L^{-1} \end{pmatrix} \,, \qquad
   g_R = \begin{pmatrix} r_R & 0 \\ 0 & r_R^{-1} \end{pmatrix} \,.
 }
Here $r_L = r_1^{(n_z+n_w)/2}$ and $r_R = r_1^{(-n_z+n_w)/2}$.  Plugging into \eno{WZWsimpler}, one sees that the relation $X_{11} = X_{10} X_{00}^{-1} X_{01}$ is preserved when passing from $X$ to $\tilde{X}$.  A peculiar point about this argument is that if $n_z$ and $n_w$ have different parity, then $r_L$ and $r_R$ will not be elements of ${\bf Z}[\sqrt{d}]$, even though $r_z$ and $r_w$ are.  In the current context, this doesn't present a problem since we can pass to a larger ring or field of numbers as an intermediate step in writing the relation \eno{idGroup}.

There are two technical points which one must confront in order to frame a fully consistent discussion of segmented strings in the BTZ background:
 \begin{enumerate}
  \item Sometimes, $SL(2,{\bf Z})_d$ has no null elements at all, meaning no elements which are null separated from ${\bf 1}$ in the standard causal structure on $SL(2,{\bf R})$.  This happens for $d=3$, and in fact for any prime $d$ congruent to $3 \mod 4$.  But for many other values of $d$, for example primes congruent to $1 \mod 4$, there are null separations in $SL(2,{\bf Z})_d$, essentially as a consequence of Fermat's two-square theorem.  In short, we must choose $d$ appropriately to ensure that $SL(2,{\bf Z})_d$ has null elements, since our description of segmented strings uses null group elements in an essential way.
  \item With an identification of points \eno{BTZid} which are spacelike separated (which is the case in regions {\bf I} and {\bf II}), there isn't a simple way to distinguish between a translation all the way around the spatial circle and no displacement at all.  This is similar to the obstacle discussed in connection with T-duality in section~\ref{TDUALITY}.  As in that case, the most straightforward resolution is to insist on working in the ``upstairs'' picture, namely the whole of regions {\bf I} and {\bf II} before identification, and restrict attention to configurations which are mapped to themselves by the identification \eno{BTZid}.  Then one may use the causal structure inherited from $SL(2,{\bf R})$, and we have a clear notion on a given serrated slice of whether a string wraps the spatial circle.  The evolution law \eno{WZWsimpler} evolves allowed serrated slices to other allowed serrated slices.  
 \end{enumerate}
Unsurprisingly, the general tendency is for strings to fall into the black hole, i.e.~to fail to remain entirely in region ${\bf I}$.  I leave as an open question the proper interpretation of string propagation that includes singular points like the $s_n$ in the case of the uncharged black hole.  Trajectories that continue into regions other than ${\bf I}$ and ${\bf II}$ can eventually explore closed timelike curves, and presumably some appropriate prescription of passing to a covering space will be needed.

\section{Conclusions}
\label{CONCLUSIONS}

The animating principle of classical segmented strings is to understand the simplest scrap of string which can be propagated forward in time in some specified background geometry.  The focus here was on triples of spacetime points $(X_{00},X_{10},X_{01})$ where $X_{10}$ and $X_{01}$ are separated from $X_{00}$ by forward-directed null displacements.  Such forward null triples, together with backward null triples $(X_{11},X_{10},X_{01})$ to which they can evolve, are a convenient starting point because we do not need to specify any momentum degrees of freedom, as we would if we considered a segment of string $(X_a,X_b)$ whose endpoints are spacelike separated.  Localized momentum is incorporated naturally into the framework of forward and backward null triples.  In a forward null triple $(X_{00},X_{10},X_{01})$, if $X_{10} = X_{01}$, then it means that the scrap of string under consideration is collapsed to a point and moving at the speed of light from $X_{00}$ to $X_{10}$.

Joining forward null triples and backward null triples together, we can construct a serrated slice of the worldsheet, and then the local rules for evolving forward null triples into backward null triples becomes a complete (classical) evolution scheme for serrated slices, which accommodates localized momentum (as at least the previous approach of \cite{Callebaut:2015fsa} did not).  An appealing feature of the null triples approach is that we do not rely upon a metric structure on spacetime.  Instead we only need an appropriate collection of forward-directed null displacements, together with a map from forward null triples to backward null triples.  This map can be thought of as a discrete replacement for the second order differential equation for the embedding of the string into spacetime.  The elementary analysis of worldsheet fermions in section~\ref{RNS} suggests that for the RNS superstring, forward and backward null triples must be augmented by information about the fermions on the legs between sites where bosonic data is located.  Possibly further augmentations could be considered that would allow for more general segmented string motions.  Another interesting possibility is to pass from real-valued coordinates to coordinates valued in a finite field.  Then segmented string evolution can be regarded as a finite-state cellular automaton.

We saw in sections~\ref{ALGEBRAIC} and \ref{WZW} that there can be more than one sensible way to propagate forward null triples into backward null triples.  In $AdS_3$, the most obvious propagation rule, following from the Nambu-Goto action, results in the algebraic evolution law \eno{Evolution}.  Because this rule involves division, it is difficult to restrict the set of allowed points in the null triples to lie in a discrete subset of $AdS_3$.  A different evolution law, \eno{WZWsimpler}, involves only group multiplication in $SL(2,{\bf R})$, which is the same as $AdS_3$ (ignoring the issue of global covers).  It is therefore easy to restrict \eno{WZWsimpler} to subgroups of $SL(2,{\bf R})$.  An interesting example is the group $SL(2,{\bf Z})_d$, as defined around \eno{gFromzw}, for square-free integers $d$.  A striking corollary is that we can define a discrete version of BTZ black holes in terms of an identification of elements of $SL(2,{\bf Z})_d$ by the action of a map \eno{BTZid} defined in terms of boosts in ${\bf R}^{2,2}$.  The entropy of the black hole, as computed by embedding it in the real manifold $AdS_3$, is then proportional to the number of points of the discrete group on the horizon.  However, the constant of proportionality involves the logarithm of the fundamental solution of the Pell equation $p^2 - d q^2 = 1$, and this logarithm varies with $d$ in an irregular fashion.  A surprising feature of the discrete BTZ black holes is that, at least for the uncharged black hole, the interior is empty until one gets to the singularity.  This is even true for black holes that are large compared to both the Planck scale and the $AdS_3$ radius.  So the firewall paradox of \cite{Almheiri:2012rt} does not arise for these discrete BTZ black holes, at least in the uncharged case.

If $AdS_3$ is supported by Neveu-Schwarz three-form flux, then the Nambu-Goto action characterizes D1-brane propagation, whereas a Wess-Zumino-Witten model characterizes fundamental string propagation.  Bound states of D1-branes and fundamental strings should lead to a whole family of evolution laws that interpolate between \eno{Evolution} and \eno{WZWsimpler}.

It is natural to inquire whether a higher dimensional version of serrated slices can be constructed.  For M2-branes, instead of forward null triples, one might try to use forward null quadruples $(A,B,C,D)$ where $B$, $C$, and $D$ are all displaced from $A$ in forward null directions: That is, they are all elements of $N^+(A)$.  Backward null quadruples can be defined similarly.  However, already in flat space one can easily see that the M2-brane worldvolume cannot be filled with unions of the convex hulls of forward and backward null quadruples.  In addition one needs tetrahedra formed as the convex hulls of quadruples $(A,B,C,D)$ where $C$ and $D$ are both in $N^+(A) \cap N^+(B)$.  I hope to report further on such constructions in the future.

\section*{Note added}

When this paper was complete, we received \cite{Vegh:2016hwq}, which includes an equation equivalent to \eno{Evolution} for the $AdS_3$ case and points out a dual relation to the evolution of normal vectors that formed the basis of \cite{Vegh:2015ska}.

\section*{Acknowledgments}

I thank P.~Landweber, S.~Parikh, P.~Witaszczyk, and A.~Yarom for useful discussions.  This work was supported in part by the Department of Energy under Grant No.~DE-FG02-91ER40671.

\bibliographystyle{ssg}
\bibliography{light}
\end{document}